\documentclass[11pt]{article}

\usepackage[a4paper,text={16cm,25.7cm},centering]{geometry}
\usepackage{bm,amssymb,amsmath,amsfonts,graphicx,mathtools}
\usepackage[numbers, comma, sort&compress]{natbib}
\usepackage{hyperref}
\usepackage[utf8]{inputenc}
\usepackage{epstopdf}
\usepackage{xcolor}

\newcommand{\beginsupplement}{%
        \setcounter{table}{0}
        \renewcommand{\thetable}{S\arabic{table}}%
        \setcounter{figure}{0}
        \renewcommand{\thefigure}{S\arabic{figure}}%
        \setcounter{equation}{0}
        \renewcommand{\theequation}{S\arabic{equation}}
     }

\begin{document}

\begin{flushleft}
{\Large
\textbf\newline{Inferring interaction partners from protein sequences using mutual information} 
}
\newline
\\
Anne-Florence Bitbol\textsuperscript{1*}
\\
\bigskip
\textbf{1} Sorbonne Université, CNRS, Laboratoire Jean Perrin (UMR 8237), F-75005 Paris, France
\\
\bigskip

* anne-florence.bitbol@sorbonne-universite.fr

\end{flushleft}
\section*{Abstract}
Functional protein-protein interactions are crucial in most cellular processes. They enable multi-protein complexes to assemble and to remain stable, and they allow signal transduction in various pathways. Functional interactions between proteins result in coevolution between the interacting partners, and thus in correlations between their sequences. Pairwise maximum-entropy based models have enabled successful inference of pairs of amino-acid residues that are in contact in the three-dimensional structure of multi-protein complexes, starting from the correlations in the sequence data of known interaction partners. Recently, algorithms inspired by these methods have been developed to identify which proteins are functional interaction partners among the paralogous proteins of two families, starting from sequence data alone. Here, we demonstrate that a slightly higher performance for partner identification can be reached by an approximate maximization of the mutual information between the sequence alignments of the two protein families. Our mutual information-based method also provides signatures of the existence of interactions between protein families. These results stand in contrast with structure prediction of proteins and of multi-protein complexes from sequence data, where pairwise maximum-entropy based global statistical models substantially improve performance compared to mutual information. Our findings entail that the statistical dependences allowing interaction partner prediction from sequence data are not restricted to the residue pairs that are in direct contact at the interface between the partner proteins.

\section*{Author summary}

Functional protein-protein interactions are at the heart of most intra-cellular processes. Mapping these interactions is thus crucial to a systems-level understanding of cells, and has broad applications to areas such as drug targeting. Systematic experimental identification of protein interaction partners is still challenging. However, a large and rapidly growing amount of sequence data is now available. Recently, algorithms have been proposed to identify which proteins interact from their sequences alone, thanks to the co-variation of the sequences of interacting proteins. These algorithms build upon inference methods that have been used with success to predict the three-dimensional structures of proteins and multi-protein complexes, and their focus is on the amino-acid residues that are in direct contact. Here, we propose a simpler method to identify which proteins interact among the paralogous proteins of two families, starting from their sequences alone. Our method relies on an approximate maximization of mutual information between the sequences of the two families, without specifically emphasizing the contacting residue pairs. We demonstrate that this method slightly outperforms the earlier one. This result highlights that partner prediction does not only rely on the identities and interactions of directly contacting amino-acids.

\section*{Keywords}
Protein-protein interactions; mutual information; inference; sequence data; information theory; paralogs; homologs


\section*{Introduction}
Most cellular processes are carried out by interacting proteins. Functional protein-protein interactions allow multi-protein complexes to assemble, and ensure proper signal transduction in various pathways. Hence, mapping functional protein-protein interactions is an important fundamental question. High-throughput experiments have recently elucidated a substantial fraction of protein-protein interactions in a few model organisms~\cite{Rajagopala14}, but such experiments remain challenging. An attractive alternative is to exploit the increasingly abundant sequence data in order to identify functional protein-protein interaction partners.

The sequences of interacting proteins are correlated, both because of evolutionary constraints arising from the need to maintain physico-chemical complementarity among amino-acids in contact, and because of shared evolutionary history. The first type of correlations has received substantial interest, both within single proteins and across protein partners, as evolutionary constraints induce correlations between amino acids that are in contact in the folded protein or in the multi-protein complex. Hence, the correlations observed in multiple sequence alignments of homologous proteins contain information about protein structure. Global statistical models allow direct and indirect correlations to be disentangled~\cite{Lapedes99,Burger08,Weigt09}. Such models, built using the maximum entropy principle~\cite{Jaynes57}, and assuming pairwise interactions, known in the field of proteins as Direct Coupling Analysis (DCA), have been used with success to determine three-dimensional protein structures from sequences~\cite{Marks11,Sulkowska12}, to analyze mutational effects~\cite{Dwyer13,Cheng14,Cheng16,Figliuzzi16} and conformational changes~\cite{Morcos11,Malinverni15}, to find residue contacts between known interaction partners~\cite{Weigt09,Procaccini11,Baldassi14,Ovchinnikov14,Hopf14,Tamir14,dosSantos15,Feinauer16}, and most recently to predict interaction partners from sequence data~\cite{Bitbol16,Gueudre16}. DCA models lay the emphasis on interactions between residues that are in direct contact in the three-dimensional protein structure. However, correlations in protein sequences also have important collective modes~\cite{Halabi09,Rivoire16}, which can arise from functional selection~\cite{Yan17,WangXX}, and additional correlations are due to phylogeny~\cite{Casari95,Halabi09,Qin18}. These contributions are deleterious to the prediction of contacts~\cite{Qin18} but not necessarily to the prediction of interacting partners, since a pair of interacting partners may be subject to common functional selection, and may also have a more strongly shared phylogenetic history than non-interacting proteins.

Here, we present an alternative approach to predict interaction partners from sequence data, among the paralogous proteins belonging to two interacting families. In contrast to the previous pairwise maximum entropy-based approaches~\cite{Bitbol16,Gueudre16}, the present method is based on an approximate maximization of mutual information between the sequences from the two protein families. Specifically, we develop a variant of the iterative pairing algorithm (IPA) introduced in~\cite{Bitbol16}, where we use mutual information (MI) as a score to maximize, instead of the effective interaction energy from a pairwise maximum entropy (DCA) model. We demonstrate that this mutual information-based algorithm (MI-IPA) performs slightly better than the one (DCA-IPA) introduced by us and colleagues in~\cite{Bitbol16}. Our findings entail that the statistical dependences allowing interaction partner prediction from sequence data are not restricted to the contacting residue pairs revealed by DCA.

\section*{Results}

We developed an iterative pairing algorithm (MI-IPA) that pairs paralogous proteins from two interacting protein families $\mathcal{A}$ and $\mathcal{B}$ by approximately maximizing mutual information between the sequences of the two families. Here, we first introduce the information theory-based pairing score we employ, before briefly explaining the steps of the MI-IPA. Next, we present the results we obtained with the MI-IPA. Throughout, we compare the performance of the MI-IPA to that obtained with the DCA-IPA from~\cite{Bitbol16}, which infers a pairwise maximum entropy model and approximately maximizes the resulting effective interaction energies. First, we consider the case where the MI-IPA starts with a training set of known protein pairs, obtaining good performance even with few training pairs. Then, we demonstrate that the MI-IPA can make accurate predictions starting without any training set, as would be needed to predict novel protein-protein interactions. Next, we assess to what extent the MI-IPA is successful at maximizing mutual information. We further demonstrate the robustness of the MI-IPA by successfully applying it to several pairs of proteins. Finally, we show how the MI-IPA reveals signatures of protein-protein interactions between two protein families. 

\subsection*{A pairing score based on pointwise mutual information (PMI)}

Consider an alignment of $M$ concatenated sequences AB of length $L$, where A is a protein from family $\mathcal{A}$ and B is a protein from family $\mathcal{B}$. At each amino-acid site $i\in\{1,..,L\}$, a given sequence can feature any amino acid (represented by $\alpha\in\{1,..,20\}$), or a gap (represented by $\alpha=21$). To describe the statistics of this concatenated alignment (CA), we employ the single-site frequencies of occurrence of each state $\alpha$ at each site $i$, denoted by $f_i(\alpha)$, and the two-site frequencies of occurrence of each ordered pair of states $(\alpha,\beta)$ at each ordered pair of sites $(i,j)$, denoted by $f_{ij}(\alpha,\beta)$. These empirical frequencies are obtained by counting the sequences where given residues occur at given sites and dividing by the number $M$ of sequences in the CA. (Note that when computing frequencies from real protein data, it is useful to weight sequences so as to attenuate the impact of biased sampling, and to include pseudocounts, in order to mitigate finite-sample effects, see Methods.) The empirical frequencies constitute estimates of the corresponding probabilities $p_i(\alpha)$ and $p_{ij}(\alpha,\beta)$, and tend toward them in the limit where the number $M$ of sequences in the alignment tends to infinity. 

The pointwise mutual information (PMI) of a pair of residues $(\alpha,\beta)$ at a pair of sites $(i,j)$ is defined as~\cite{Fano61}:
\begin{equation}
\textrm{PMI}_{ij}(\alpha,\beta)=\log\left[\frac{p_{ij}(\alpha,\beta)}{p_i(\alpha)p_j(\beta)}\right]\underset{M\gg 1}\approx\log\left[\frac{f_{ij}(\alpha,\beta)}{f_i(\alpha)f_j(\beta)}\right]\,.
\label{PMI}
\end{equation}
Averaging this quantity over all possible residue pairs yields the mutual information (MI) between sites $i$ and $j$~\cite{Cover06}:
\begin{equation}
\textrm{MI}_{ij}=\sum_{\alpha,\beta}p_{ij}(\alpha,\beta)\,\textrm{PMI}_{ij}(\alpha,\beta)=\sum_{\alpha,\beta}p_{ij}(\alpha,\beta)\log\left[\frac{p_{ij}(\alpha,\beta)}{p_i(\alpha)p_j(\beta)}\right]\,.
\label{MI}
\end{equation}
PMI has been used in linguistics to study the co-occurrence of words~\cite{Church90,Role11}. Note that in some instances~\cite{Fano61,Church90} PMI is called MI, and MI is then referred to as the average value of MI.

We define a pairing score $S_\mathrm{AB}$ for each pair AB of proteins as the sum of the PMIs of the inter-protein pairs of sites of this concatenated sequence (i.e. those that involve one site in protein A and one site in protein B):
\begin{equation}
S_\mathrm{AB}=\sum_{i=1}^{L_\mathrm{A}}\sum_{j=L_\mathrm{A}+1}^{L}\textrm{PMI}_{ij}(\alpha_i,\beta_j)\,,
\label{SAB}
\end{equation}
where we have denoted the concatenated sequence AB by $(\alpha_1,\dots,\alpha_{L_\mathrm{A}},\beta_{L_\mathrm{A}+1},\cdots,\beta_L)$, with $L_\mathrm{A}$ the length of the A sequence. This score can be computed for a pair AB that is a member of the CA used to estimate the PMI of each residue pair at each site, but also for any other pair AB comprised of the sequences of members of the protein families $\mathcal{A}$ and $\mathcal{B}$.

Next, consider a candidate assignment X of $M'$ pairs AB, where each protein A is paired with a protein B from the same species, resulting in a CA of $M'$ sequences of length $L=L_\mathrm{A}+L_\mathrm{B}$. Again, the pairs in this CA can involve the proteins in the CA of $M$ pairs used to estimate the PMIs, with the same assignment or a different one, or any other pair AB comprised of the sequences of members of the protein families $\mathcal{A}$ and $\mathcal{B}$. We define the overall pairing score $S_\mathrm{X}$ of the assignment $X$ by the average of all pairing scores (see Eq.~\ref{SAB}) of the pairs involved:
\begin{align}
S_\mathrm{X}&=\frac{1}{M'}\sum_{\mathrm{AB}\in\mathrm{X}}S_\mathrm{AB}=\frac{1}{M'}\sum_{i=1}^{L_\mathrm{A}}\sum_{j=L_\mathrm{A}+1}^{L}\sum_{\mathrm{AB}\in\mathrm{X}}\textrm{PMI}_{ij}(\alpha_i,\beta_j)\nonumber\\
&=\frac{1}{M'}\sum_{i=1}^{L_\mathrm{A}}\sum_{j=L_\mathrm{A}+1}^{L}\sum_{\alpha,\beta}\sum_{\substack{\mathrm{AB}\in\mathrm{X}\\
                              \alpha_i=\alpha,\,\beta_j=\beta}}\textrm{PMI}_{ij}(\alpha,\beta)=\sum_{i=1}^{L_\mathrm{A}}\sum_{j=L_\mathrm{A}+1}^{L}\sum_{\alpha,\beta}f'_{ij}(\alpha,\beta)\,\textrm{PMI}_{ij}(\alpha,\beta)
\,,
\label{Sx1}
\end{align}

where $f'_{ij}(\alpha,\beta)$ denotes the joint empirical frequency of amino acid state $\alpha$ at site $i$ and amino acid state $\beta$ at site $j$ in X. It corresponds to the number of concatenated sequences in X featuring both amino acid state $\alpha$ at site $i$ and amino acid state $\beta$ at site $j$, divided by $M'$.

In the limit of large alignments, the empirical frequencies tend toward probabilities. Besides, if the scored CA is the same as the one used to calculate the PMIs, all frequencies will be the same for both of them (i.e. $f'_{ij}(\alpha,\beta)=f_{ij}(\alpha,\beta)$). For different CA of proteins from the same families $\mathcal{A}$ and $\mathcal{B}$, in the case of an assignment consistent with the CA used to estimate the PMIs, the limiting two-body probabilities will be the same in the two CA. Hence, in these cases, combining Eqs.~\ref{MI} and~\ref{Sx1} yields
\begin{equation}
S_\mathrm{X}\xrightarrow[\substack{M\to \infty\\M'\to \infty}]{} \sum_{i=1}^{L_\mathrm{A}}\sum_{j=L_\mathrm{A}+1}^{L}\mathrm{MI}_{ij}\,.
\label{MIlim}
\end{equation}

For large alignments, maximizing $S_\mathrm{X}$ thus corresponds to maximizing the sum of the MIs of inter-protein site pairs, which is itself a pairwise approximation of the MI between the sequences of two protein families $\mathcal{A}$ and $\mathcal{B}$. A brute-force self-consistent maximization of $S_\mathrm{X}$ over all possible assignments X of a realistic dataset would result in a combinatorial explosion, since all allowed permutations of pairs would need to be considered, and PMIs would need to be computed for each of them. In practice, since biologically meaningful pairings have to be made within a species, this means that we would need to consider all combinations of all permutations within each species, which already yields prohibitively large numbers of assignments to test. Hence, we propose an algorithm to perform an approximate maximization of $S_\mathrm{X}$.

\subsection*{An iterative pairing algorithm (IPA) based on MI}

In order to approximately maximize mutual information via the score $S_\mathrm{X}$ (see Eq.~\ref{Sx1}), we propose an iterative pairing algorithm (referred to as MI-IPA) inspired by that of~\cite{Bitbol16} (referred to as DCA-IPA), where the effective interaction energy from a global statistical model was approximately maximized. 

In each iteration, we first estimate PMIs for all inter-protein residue pairs from a concatenated alignment (CA) of paired sequences. The initial CA, used at the first iteration, is either built from a training set of known correct protein pairs, or made from random pairs, assuming no prior knowledge of interacting pairs. We calculate the pairing scores $S_\mathrm{AB}$ (see Eq.~\ref{SAB}) for every possible protein pair AB within each species, by summing the inter-protein PMIs. Next, within each species, we assign pairs by maximizing the sum of $S_\mathrm{AB}$ scores in the species (assuming one-to-one specific interactions), thereby maximizing $S_\mathrm{X}$ (see Eq.~\ref{Sx1}) over biologically relevant pair assignments, where each protein has a partner within its species. We attribute a confidence score to each predicted pair, by using the difference of scores between the optimal assignment of pairs in the species and the best alternative assignment that does not involve this predicted pair. The CA is then updated by including the highest-scoring protein pairs, and the next iteration can begin. At each iteration, all pairs in the CA are re-selected based on confidence scores (except the initial training pairs, if any), allowing for error correction. More details on each step of the MI-IPA are given in Methods.

\subsection*{The MI-IPA accurately predicts interaction partners from a training set of known partners}

As in~\cite{Bitbol16}, we use histidine kinases (HKs) and response regulators (RRs) from prokaryotic two-component signaling systems as our main benchmark. Two-component systems are important pathways that enable bacteria to sense and respond to environment signals. Typically, a transmembrane HK senses a signal, autophosphorylates, and transfers its phosphate group to its cognate RR, which in turn induces a cellular response~\cite{Laub07}. Importantly, most cognate HK-RR pairs are encoded in the same operon, so actual interaction partners are known, which enables us to assess performance.

Unless otherwise specified, our results were obtained on a ``standard dataset'' comprising 5064 HK-RR pairs for which the correct pairings are known from gene adjacency. Each species has on average $\langle m_p\rangle=11.0$ pairs, and at least two pairs (see Methods).

We start by predicting interaction partners starting from a training set of known pairs. As our training set, we pick a random set of $N_\mathrm{start}$ known HK-RR pairs from the standard dataset. The first iteration of the MI-IPA uses this concatenated alignment (CA) to compute PMIs and score possible pairs. We blind the pairings of the remaining dataset, and use it as a testing set on which we predict pairings. At each subsequent iteration $n>1$, the CA used to recompute PMIs contains the initial training pairs plus the $(n-1) N_\mathrm{increment}$ highest-scoring predicted pairs from the previous iteration (see Methods). 

\begin{figure}[h t b]
\centering
\includegraphics[width=\textwidth]{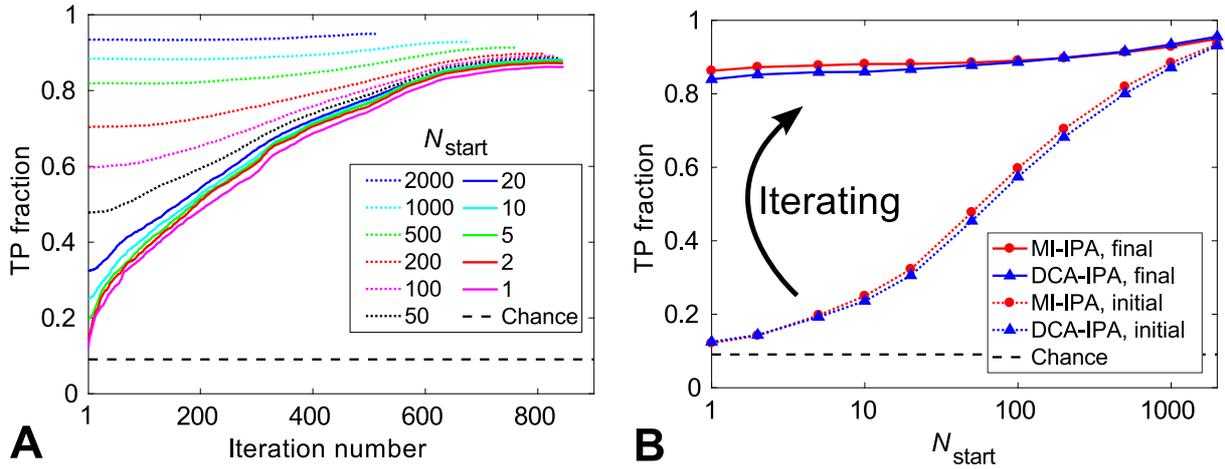}\\
\caption{\label{Fig1}\textbf{Performance of the MI-IPA for different training set sizes $N_{\mathrm{start}}$.} (A) Increase of the TP fraction with the number of iterations during the MI-IPA. (B) Initial and final TP fractions (at the first and last iteration) versus $N_{\mathrm{start}}$ for the MI-IPA and for the DCA-IPA of~\cite{Bitbol16}. In both panels, the standard dataset of HK-RRs is used, and the CA includes $N_{\mathrm{increment}}=6$ additional pairs at each iteration. All results are averaged over 50 replicates that differ by the random choice of HK-RR pairs in the training set. Dashed lines represent the average TP fraction obtained for random within-species HK-RR pairings. }
\end{figure}

As in the case of the DCA-IPA~\cite{Bitbol16}, iterating, and thereby progressively adding high-scoring pairs to the CA, allows us to increase the fraction of pairs that are correctly predicted. This gradual improvement of the TP fraction during the iterations of the MI-IPA is shown in Fig.~\ref{Fig1}A for different training set sizes $N_\mathrm{start}$. The increase of TP fraction is especially spectacular for small training sets. Fig.~\ref{Fig1}B shows the initial TP fraction, obtained at the first iteration, and the final TP fraction, obtained at the last iteration, versus the size of the training set $N_\mathrm{start}$, both for the MI-IPA and for the DCA-IPA~\cite{Bitbol16}. In both cases, comparing the initial and final TP fractions demonstrates the major interest of our iterative approach, through the massive increase in TP fraction, especially for small training sets.  Moreover, the final TP fraction depends only weakly on $N_\mathrm{start}$: the iterative approach removes the need for large training sets. Both algorithms yield very good performance, and the MI-IPA even outperforms the DCA-IPA in the trickiest case of small training sets. In this limit ($N_\mathrm{start}=1$), the MI-IPA yields 86\% true positive (TP) pairs while the DCA-IPA yields 84\% TP, while both start from 12\% TP at the first iteration. These final TP fractions are strikingly higher than the random expectation of 9\%, while such small training sets contain very little information about pairings, as illustrated by the associated low initial TP fraction.

Since the MI-IPA to the DCA-IPA yield similar performance, we asked whether they tend to predict the same correct pairs when starting from the same training set. To assess this, consider a species with $m$ AB pairs, and denote by $p$ (resp. $q$) the number of pairs correctly assigned by the MI-IPA (resp. by the DCA-IPA) in this species. If the two algorithms made independent predictions, it would correspond to independently and randomly drawing $p$ (resp. $q$) proteins A among $m$, to be correctly paired. In this null model, the number of possible MI-IPA assignments that share $k$ correct pairs with the DCA-IPA is $\binom{q}{k} \binom{m-q}{p-k}$: $k$ correct pairs are chosen among the $q$ pairs correctly assigned by the DCA-IPA, and the other $p-k$ ones are chosen among the $m-q$ proteins A incorrectly paired by the DCA-IPA. Besides, the total number of possible MI-IPA assignments with $p$ correct pairs is $\binom{m}{p}$. Hence, the probability $P(p,k,m,q)$ that $k$ correct pairs are predicted by both algorithms is given by the hypergeometric distribution:
\begin{equation}
 P(p,k,m,q)=\frac{\binom{q}{k} \binom{m-q}{p-k}}{\binom{m}{p}}\,.
\end{equation}
The expectation of $k$ under this distribution is given by 
\begin{equation}
 \langle k\rangle=\frac{p\,q}{m}\,.
 \label{expect}
\end{equation}
Hence, in this fully independent null model, the expectation of the total number of correct pairs predicted by both algorithms can be obtained by summing the expectations in Eq.~\ref{expect} over all species in the dataset. Moreover, in each species, the observed number $k_\mathrm{obs}$ of shared correct pairs can be compared to this null model, as well as to the extreme case where all correct pairs that can be shared are shared, yielding $\min(p,q)$ shared pairs. Hence, we define the relative excess $E$ of shared predictions by
\begin{equation}
E=\frac{\sum_{i=1}^S k_{\mathrm{obs},i}-\langle k_i \rangle}{\sum_{i=1}^S\min(p_i,q_i)-\langle k_i \rangle}\,,
\end{equation}
where the index $i$ corresponds to a particular species, the sums run over the $S$ species present in the dataset, and the expectations are given by Eq.~\ref{expect}. If $E>0$, the two algorithms tend to predict the same pairs more frequently than if their predictions were fully independent. In addition, the maximal value $E$ can take is 1, including the case where predictions from both algorithms are exactly the same. We calculated the average value of $E$ across 50 replicates where both algorithms were started from the same training set, for various $N_\mathrm{start}$ values, in the same conditions as in Fig.~\ref{Fig1}. As expected, we found that $E$ increases with $N_\mathrm{start}$, as the algorithms share more information to begin with. However, $E$ depends rather weakly on the size $N_\mathrm{start}$ of the training set, varying smoothly from 53\% for $N_\mathrm{start}=1$ to 68\% for  $N_\mathrm{start}=2000$. This indicates a significant tendency of the two algorithms to make the same correct predictions, even in the case of small training sets. 

\subsection*{Great accuracy is maintained in the absence of a training set}

In~\cite{Bitbol16}, we showed that the DCA-IPA yields very good identification of interacting pairs without any training set, i.e. without any prior knowledge of interacting pairs. Given this previous result, and given the success of the MI-IPA with very small training sets, we ask whether the MI-IPA also makes good predictions in the absence of a training set. To test this, we followed the approach introduced in~\cite{Bitbol16} by randomly pairing each HK with an RR from the same species, and using these 5064 random pairs to train the initial model. At each subsequent iteration $n > 1$, the CA only contained the $(n-1)N_\mathrm{increment}$ highest-scoring pairs from the previous iteration (see Methods).

\begin{figure}[h t b]
\centering
\includegraphics[width=\textwidth]{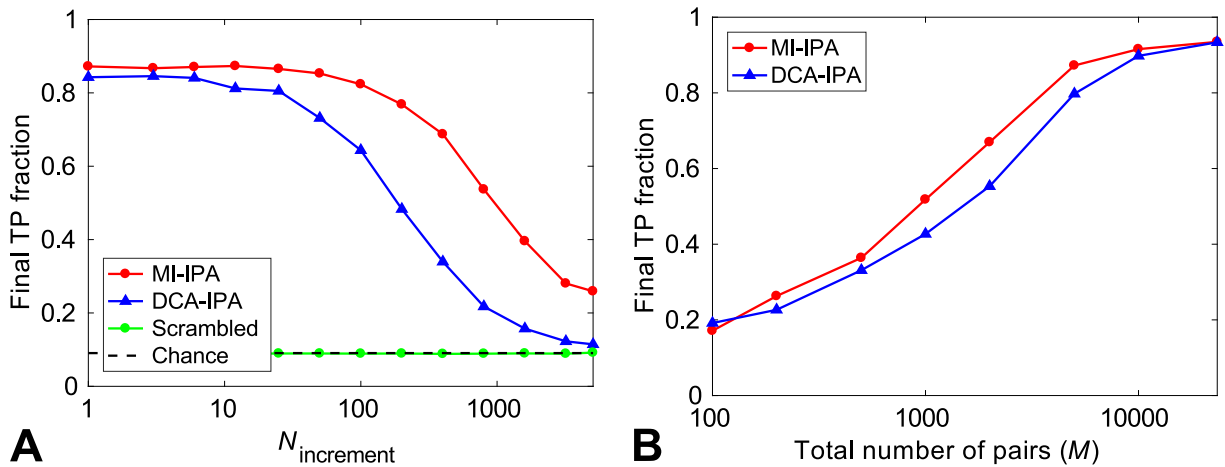}\\
\caption{\label{Fig2}\textbf{Performance of the MI-IPA without a training set.} (A) Final TP fraction obtained by the MI-IPA and the DCA-IPA versus $N_{\mathrm{increment}}$ for the standard HK-RR dataset. At the first iteration, the CA is made of random within-species HK-RR pairs. At each subsequent iteration $n>1$, the CA includes the $(n-1)N_{\mathrm{increment}}$ top predicted pairs. All results are averaged over 50 replicates employing different initial random pairings. The MI-IPA was also run on an alignment where each column is randomly scrambled. The dashed line represents the average TP fraction obtained for random within-species HK-RR pairings. (B) Final TP fraction obtained by the MI-IPA and the DCA-IPA versus the total number $M$ of HK-RR pairs in the dataset, starting from random pairings. For each $M$, except that corresponding to the full dataset, datasets are constructed by picking species randomly from the full dataset, and results are averaged over multiple different such alignments (from 50 up to 500 for small $M$). For the full dataset (largest $M$), averaging is done on 50 different initial random within-species pairings. All results in (B) are obtained in the small-$N_{\mathrm{increment}}$ limit. }
\end{figure}

Fig.~\ref{Fig2}A shows the final TP fraction obtained for different values of $N_\mathrm{increment}$, both for the MI-IPA and for the DCA-IPA. In both cases, the iterative method performs best for small increment steps, which highlights again the interest of the iterative approach. Importantly, the MI-IPA performs better than the DCA-IPA for all values of $N_\mathrm{increment}$, and requires substantially less small increments than the DCA-IPA to reach the same performance. The low--$N_\mathrm{increment}$ limit of the final TP fraction is 0.87 for the MI-IPA, versus 0.84 for the DCA-IPA. These values are consistent with those obtained above with a single training pair, $N_\mathrm{start}=1$ (Fig.~\ref{Fig1}A). We emphasize that the striking TP fraction of 0.87 is attained by the MI-IPA without any prior knowledge of HK-RR interactions. Ref.~\cite{Bitbol16} showed that an important ingredient for the DCA-IPA to bootstrap its way toward high predictive power is that sequence similarity is favored at early iterations, which increases the TP fraction in the CA, because correct pairs have more neighbors in terms of Hamming distance than incorrect pairs. In~\cite{Bitbol16}, this was called the \textit{Anna Karenina effect}, in reference to the first sentence of Tolstoy's novel. The same explanation holds for the success of the MI-IPA starting from no training set. In addition, both with MI and with DCA, when starting from random pairings, signal from actual partners should add constructively, while noise should add incoherently. To confirm that the predictive power arises from correlations and similarities in the data, we ran the MI-IPA on a version of our standard alignment of HK-RR sequences where amino acids at each site (each column) are randomly scrambled, thus removing correlations (shown in green in Fig.~\ref{Fig1}A). As expected, performance is then the same as for random within-species pairings (dashed line in Fig.~\ref{Fig1}A).

In Fig.~\ref{FigS1}A, we investigated the performance of other variants of the IPA in the same conditions as in Fig.~\ref{Fig2}A. Instead of PMIs, we explored the possibilities of using normalized PMIs (NPMIs), $\mathrm{NPMI}_{ij}(\alpha,\beta)=-\mathrm{PMI}_{ij}(\alpha,\beta)/\log(f_{ij}(\alpha,\beta))$~\cite{Role11} and covariances, $C_{ij}(\alpha,\beta)=f_{ij}(\alpha,\beta)-f_{i}(\alpha)f_{j}(\beta)$ (recall that in the mean-field approximation, DCA employs the inverse of this covariance matrix). For small $N_\mathrm{increment}$ values, we find that NPMIs perform similarly as PMIs, while covariances do significantly worse than both PMIs and DCA scores.

An important parameter for the performance of the MI-IPA is dataset size. Qualitatively, larger datasets imply more close neighbors, which is favorable to the success of bootstrapping, and they also allow one to estimate MI more accurately, so we expect the MI-IPA to perform best for large datasets. Indeed, Fig.~\ref{Fig2}B shows that the performance of the MI-IPA increases with dataset size. Moreover, the rise of performance occurs for slightly smaller datasets in the case of the MI-IPA than for the DCA-IPA.  With DCA, a sufficiently large dataset is necessary to properly infer the pairwise maximum-entropy based global statistical model at the heart of the method~\cite{Weigt09, Marks11,Morcos11}. While being data-thirsty too, the MI-IPA bypasses this particular need, and thus, it is better suited for partner prediction in smaller datasets. For the complete dataset (23,424 HK-RR pairs, see Methods), both methods reach the same striking final TP fraction of 0.93. 

The fact that the MI-IPA often performs better than the DCA-IPA for predicting interacting partners stands in contrast with the fact that DCA substantially outperforms MI for residue contact prediction, and suggests that relevant covariation information is contained in pairs of residues that are not in contact. In order to test this, we ran the MI-IPA starting from random initial pairings on our standard HK-RR dataset, but suppressing all MI contributions from the contacting pairs of residues (defined with a generous $8\,\text{\AA}$ threshold on the minimum distance between amino acids). For $N_\mathrm{increment}=6$, a striking TP fraction of 0.83 was obtained. This value is only slightly smaller than the 0.87 TP fraction obtained when all residue pairs are included, thus confirming that substantial MI relevant for partnership prediction is present in non-contacting residue pairs. 

As above, we investigated the extent to which the MI-IPA and the DCA-IPA tend to make the same correct predictions. Note that this time, the two algorithms do not start from shared information since there is no training set. For the standard dataset of 5064 HK-RR pairs, with $N_\mathrm{increment}=6$, we found that the average excess shared fraction (see Eq.~\ref{expect}) of correct pairs between the two algorithms is $E=53\%$, i.e. the same as the value obtained in the limit of a very small shared training set, $N_\mathrm{start}=1$ (see above). Recall that this positive value means that the two algorithms tend to make the same correct predictions. Moreover, the bias is 53\% of the maximal value it could take. It is interesting to compare this result to the bias toward shared correct predictions across different replicates of the same algorithm. We obtained $E=73\%$ for the DCA-IPA, and $E=94\%$ for the MI-IPA, meaning that predictions from the two different algorithms are less similar than those made by the same algorithm. 

Since the two algorithms do make some different correct predictions, we next asked whether this can exploited, with the intuition that pairs that are predicted both by the MI score and by the DCA score will tend to be correct more often than other pairs. First, we considered using both scores within the IPA. We tried a combined IPA which calculates both scores at each iteration and computes separately the two corresponding pair assignments. Our ranking of pairs puts first the pairs contained in both assignments, and these pairs are ordered by decreasing MI-based confidence scores. Next come the other pairs: the assignment predicted using MI is conserved for them, and they are ordered by decreasing MI-based confidence scores. This means that the pairs consistently predicted by both methods are going to enter the CA earlier in the iterative process than those that differ. A very minor improvement was obtained over the MI-IPA using this method (see Fig.~\ref{FigS1}B). Since DCA typically requires large datasets to be reliable, we also tried combining MI and DCA by using MI at early iterations, when the CA is small, and switching to DCA at later iterations. No improvement over the MI-IPA was obtained using this method (see Fig.~\ref{FigS1}B). 

Next, we tried combining final results from the DCA-IPA and the MI-IPA. In~\cite{Bitbol16}, we showed that multiple different random initializations of the DCA-IPA can be exploited to increase the TP fraction. Specifically, when ranking all possible HK-RR pairs by the fraction of replicates of the MI-IPA in which they are predicted (``replication fraction''), in decreasing order, we found a TP fraction of 0.89 among the $M$ best-ranked pairs, a significant improvement over the 0.84 average of TP fractions obtained in individual replicates. (Here, $N_\mathrm{increment}=6$ is used. Recall that $M=5064$ is the total umber of pairs in the CA of HK-RRs, see Methods, and thus the number of predicted pairs in each individual replicate of the algorithm.) Applying the same strategy to the MI-IPA also yields a TP fraction of 0.89, higher than that obtained from individual replicates, 0.87. In both cases, 500 replicates of the algorithms, differing only by their initial random within-species pairings, were used to estimate replication fractions. Among the $M$ top pairs thus predicted by the two separate methods, 87\% were common, and the TP fraction among those common pairs was 0.97. We also combined the results from the 500 replicates of the DCA-IPA and the 500 replicates of the MI-IPA, and ranked pairs using their overall replication fraction. Then the TP fraction among the $M$ best-ranked pairs is 0.91. Hence, combining final predictions from both methods yields a further improvement of performance.

\subsection*{The MI-IPA reaches near-maximal MI}

The MI-IPA approximately maximizes MI between the sequence alignments of two protein families $\mathcal{A}$ and $\mathcal{B}$. However, there is not guarantee that it will find the assignment with highest MI. In addition, the score we maximize converges toward the sum of MIs of all inter-protein residue pairs (henceforth called ``pairwise MI") only in the limit of large alignments and assignments consistent with the CA used to estimate the PMIs (see Eq.~\ref{MIlim}). In practice, how well does the MI-IPA approach the goal of maximizing pairwise MI?

\begin{figure}[h t b]
\centering
\includegraphics[width=\textwidth]{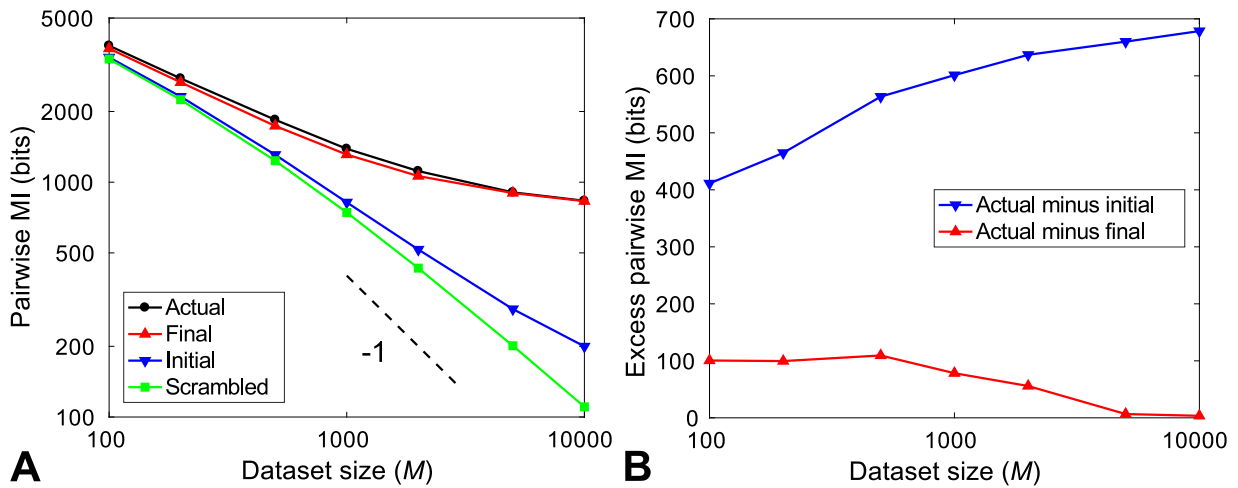}\\
\caption{\label{Fig3}\textbf{Increase of pairwise MI obtained by the MI-IPA.} (A) Estimates of the pairwise MI (sum of MIs of all inter-protein residue pairs) are shown versus the total number $M$ of sequences in the dataset. The different curves correspond to the pairwise MI of the actual set of correctly paired HK-RRs, of the initial random within-species assignment used to initialize the MI-IPA, and of the final assignment predicted by the MI-IPA, as well as of an alignment where each column was scrambled. Both axes have a logarithmic scale. The slope $-1$, expected from leading-order finite-size effects, is indicated by the dashed line. For each $M$, HK-RR datasets are constructed by picking species randomly from the full dataset, and results are averaged over 50 different such alignments. (B) Excess pairwise MI of the actual set of correctly paired HK-RRs, compared to the initial random within-species pair assignments and to the final assignment predicted by the MI-IPA. The MI-IPA successfully reduces this excess pairwise MI, thus approaching the pairwise MI of the actual alignment. Same data as in (A). }
\end{figure}

In order to answer this question, we now compare the pairwise MI of the pairs assigned by the MI-IPA to that of the actual protein pairs, and to the pairwise MI of the random within-species assignment which is used to initialize the MI-IPA in the absence of a training set. Fig.~\ref{Fig3}A shows these three quantities as a function of dataset size $M$, for HK-RR datasets to which the MI-IPA is applied starting from no training set, as in Fig.~\ref{Fig2}B. We observe a global trend of all computed pairwise MIs to decrease when $M$ is increased. This arises from a well-known finite size effect that occurs when estimating entropies from real datasets~\cite{Bialek}, and thus affects entropy-derived quantities such as MIs. To illustrate this point, Fig.~\ref{Fig3}A also shows the pairwise MI of HK-RR datasets where each column of the alignment is randomly scrambled, thus destroying actual correlations while retaining finite-size noise, as well as one-body frequencies. This null model features a similar decreasing trend as the other curves, thus demonstrating that this trend comes from finite-size effects. Note also that the scrambled alignment features pairwise MI values close to those of the initial random assignment, which makes sense because in both cases inter-protein residue pairs are decorrelated by the scrambling. The slightly higher MI of the initial random assignment arises from the fact that random partners are chosen within each species in this case, while complete columns are scrambled in the null model. Apart from the downward trend, a striking observation from Fig.~\ref{Fig3}A is that the pairwise MIs of the assignments predicted by the MI-IPA are significantly higher than those of the initial random assignments, and close to those of the actual protein pairs. Fig.~\ref{Fig3}B highlights these points by considering the excess pairwise MI in the actual protein pairs versus the initial and final assignments in the MI-IPA. Interestingly, even for small datasets, where the MI-IPA yields small TP fractions (see Fig.~\ref{Fig2}B), the pairwise MI of the assignment predicted by the MI-IPA is much closer to that of the actual assignment than to that of the initial random assignment. This suggests that the MI-IPA does a good job at maximizing pairwise MI, even though it does not reach the absolute maximum.

So far, we have used naive estimates of the pairwise MI, employing empirical frequencies instead of probabilities in MI (see Eq.~\ref{MI}), without correcting for the finite-size effect discussed above and visible in Fig.~\ref{Fig3}A. Various approaches have been proposed in order to correct for the systematic error due to finite-size effects in entropy (and thus MI) estimates. One can use the fact that these finite-size effects have a leading-order contribution of the form $\widehat{\textrm{MI}}-\textrm{MI}\propto K/M$, where $\textrm{MI}$ is the actual value of the mutual information and $\widehat{\textrm{MI}}$ is its naive estimate using empirical frequencies, while $K$ is the number of independent values that can be taken by the pair of random variables considered, and $M$ represents dataset size~\cite{Bialek}. Estimates $\widehat{\textrm{MI}}$ obtained by subsampling the initial dataset can be linearly (or polynomially, to include higher-order corrections) fitted versus $1/M$, yielding the actual MI as the intercept. However, if $M$ is not larger than $K$, subleading corrections become too important, and this method becomes inaccurate. Here, $K$ can go up to the square of the number of possible states per site, i.e. $\sim\!400$. Consistently, we found that polynomial fits become poor when considering datasets with less than $\sim\!1000$ sequences. Fig.~\ref{FigS2}A presents results obtained using corrections from these fits, for datasets of at least 1000 sequences. A more sophisticated method to reliably estimating the entropies of discrete distributions was introduced in~\cite{Nemenman02}. It employs a Bayesian approach and a flat prior on the entropy, and results in a correction of finite-size bias in entropy estimates. This NSB (Nemenman-Shafee-Bialek) correction is successful in undersampled cases~\cite{Nemenman02,Nemenman04}. In Fig.~\ref{FigS2}B, we show results obtained on our datasets using the NSB correction. Employing either of these two corrections of finite-size effects on the MI estimates (see Fig.~\ref{FigS2}) confirms our previous conclusion. For all dataset sizes, the pairwise MI of the assignment predicted by the MI-IPA is significantly closer to that of the actual assignment than to that of the initial random assignment, thus confirming that the MI-IPA yields results with near-maximal pairwise MI.

\subsection*{Performance of the MI-IPA is robust across various protein pairs}

To demonstrate the generality of the MI-IPA, we applied it to several pairs of protein families~\cite{Ovchinnikov14}, beyond HK-RRs. First, we considered several protein pairs involved in ABC transporter complexes, which permit the translocation of different substances across cell membranes~\cite{Rees09}. We built alignments of homologs of the \textit{Escherichia coli} interacting protein pairs MALG-MALK, FBPB-FBPC, and GSIC-GSID, all involved in ABC transporter complexes, using the same method as in~\cite{Bitbol16} (see Methods). As in the case of HK-RRs, for each of these pairs of protein families, we worked on subsets of $\sim 5000$ protein pairs from species containing at least two pairs. In addition, we considered smaller families of proteins, yielding $\sim 2000$ pairs. More specifically, we chose two pairs of proteins involved in enzymatic complexes: PAAH-PAAJ is a pair of proteins involved in the fatty acid $\beta$-oxidation multienzyme complex~\cite{Ishikawa04}, and XDHC-XDHA is a pair of proteins involved in the xanthine dehydrogenase complex~\cite{Dietzel09}.

\begin{figure}[h t b]
\centering
\includegraphics[width=\textwidth]{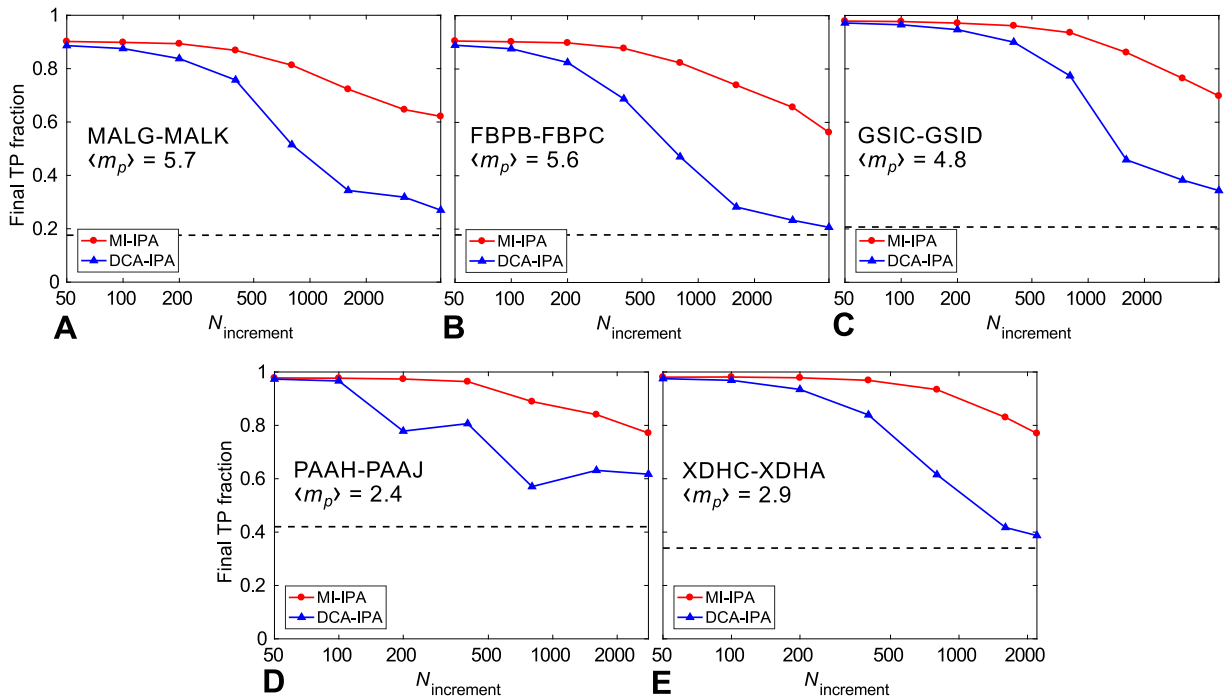}\\
\caption{\label{Fig4}\textbf{Performance of the MI-IPA for different protein pairs.} Final TP fraction obtained without a training set by the MI-IPA and the DCA-IPA versus $N_{\mathrm{increment}}$. (A,B,C) Pairs involved in ABC transporters; datasets of $\sim 5000$ pairs extracted from larger full datasets. (D,E) Pairs involved in enzymatic complexes; full datasets of $\sim 2000$ pairs. In each case, the mean number $\langle m_p\rangle$ of pairs per species is indicated, and dashed lines represent the average TP fraction obtained for random within-species pairings. All results are averaged over 50 replicates that differ in their initial random within-species pairings. }
\end{figure}

Fig.~\ref{Fig4} shows the performance of the MI-IPA and of the DCA-IPA for these protein pairs, starting from no initial training set, for various values of $N_\mathrm{increment}$. In all cases, we find that both methods perform very well for small $N_\mathrm{increment}$, but that higher performances are reached by the MI-IPA, particularly at larger $N_\mathrm{increment}$ values. This is consistent with the results obtained with HK-RRs (see Fig.~\ref{Fig2}A). Note that here, compared to HK-RRs, larger $N_\mathrm{increment}$ values suffice to obtain good performance. This is due to the fact that the pairing tasks are easier here, since the average number of proteins pairs per species is smaller (see Fig.~\ref{Fig4}, to be compared to $\langle m_p\rangle=11.0$ for HK-RRs).

These accurate predictions demonstrate the broad applicability of the MI-IPA. Note that HK-RRs interact transiently, while the ABC transporter and enzymatic proteins we considered form permanent complexes, which highlights the generality of the method.

\subsection*{Predicting protein-protein interactions}

\begin{figure}[h t b]
\centering
\includegraphics[width=\textwidth]{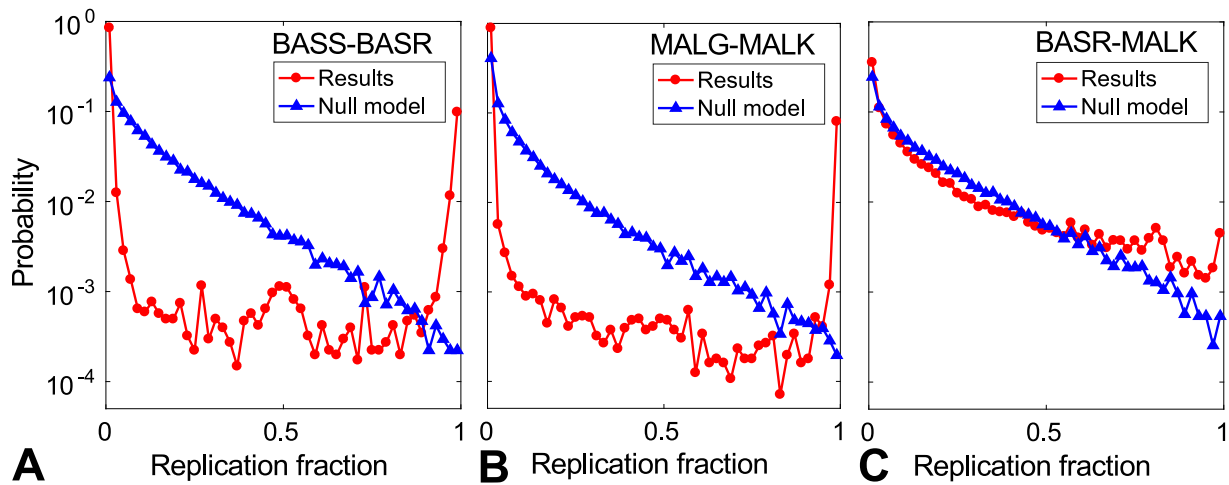}\\
\caption{\label{Fig5}\textbf{Signature of protein-protein interactions. } Distribution of the fraction of MI-IPA replicates where each possible within-species protein pair is predicted as a pair, for three different pairs of protein families. The MI-IPA replicates only differ in their initial random pairings. (A and B) Interacting protein families: BASS-BASR homologs and MALG-MALK homologs. (C) Protein families with no known interaction (BASR-MALK homologs). In all panels, the distribution of replication fraction obtained on actual sequence alignments is shown together with the same distribution obtained by running the MI-IPA on alignments where each column is randomly scrambled (null model). All alignments include $\sim\!5000$ protein pairs, with average number of pairs per species $\langle m_p\rangle\approx 5$, and each distribution is estimated from 500 MI-IPA replicates, using $N_\mathrm{increment}=50$. }
\end{figure}

\begin{figure}[h t b]
\centering
\includegraphics[width=\textwidth]{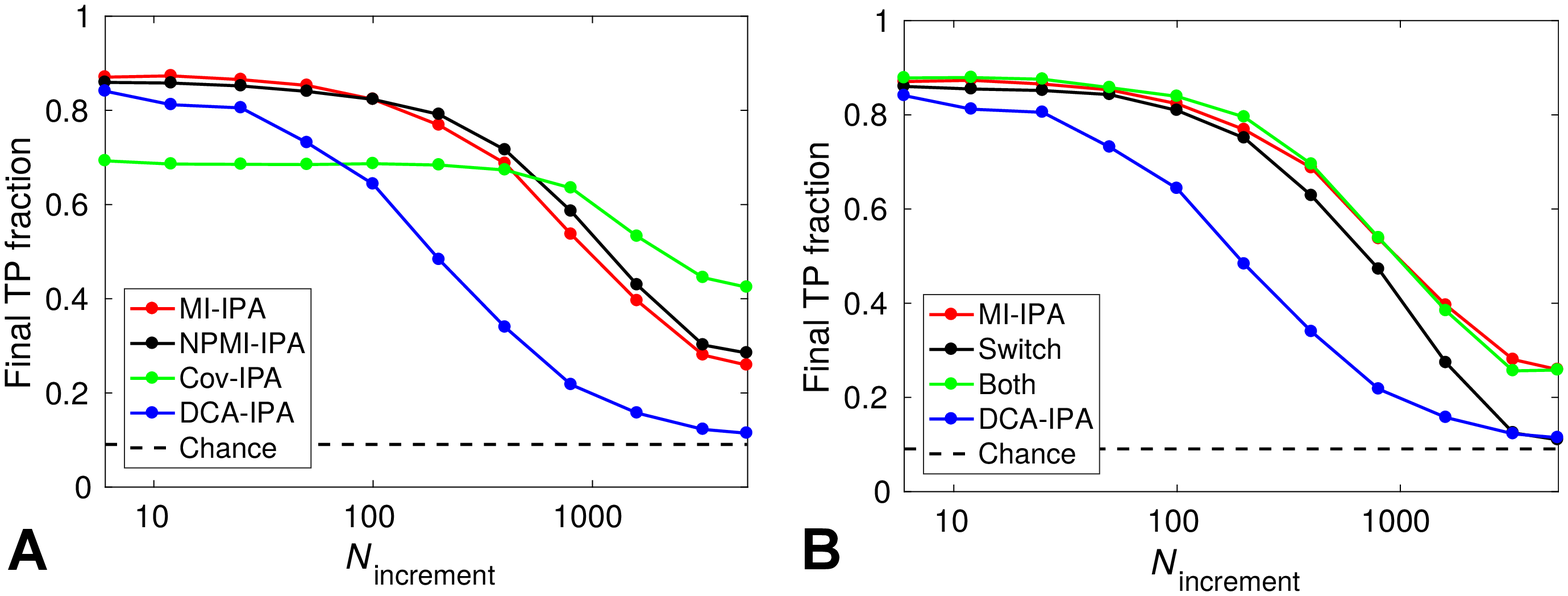}\\
\caption{\label{FigS1}\textbf{IPA variants.} (A) Final TP fraction obtained versus $N_{\mathrm{increment}}$ for the standard HK-RR dataset, starting from random within-species HK-RR pairs. In addition to the MI-IPA and the DCA-IPA (see Fig.~\ref{Fig2}A), two other variants of the IPA are presented, which score inter-protein residue pairs by the normalized PMI (NPMI-IPA) and by the covariance (Cov-IPA), respectively (see Results). Apart from this scoring difference, all particulars are the same as in the MI-IPA (see Methods). (B) Similar graph as in (A), showing two variants of the IPA that combine MI and DCA: ``Switch'' uses the MI-IPA for the first half of iterations, before switching to the DCA-IPA; ``Both'' uses both MI and DCA at each iteration and favors protein pairs that are predicted by both methods (see Results). In both panels, all results are averaged over 50 replicates employing different initial random pairings, and the dashed line represents the average TP fraction obtained for random within-species HK-RR pairings.}
\end{figure}

\begin{figure}[h t b]
\centering
\includegraphics[width=\textwidth]{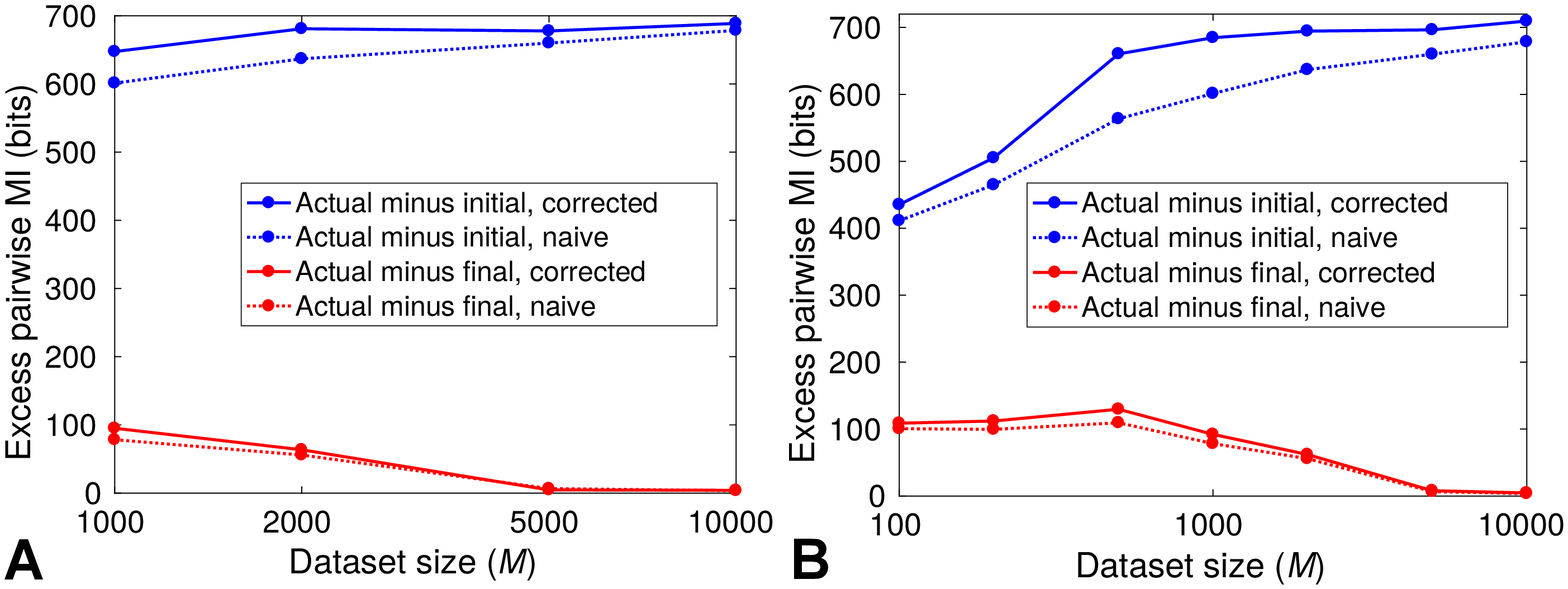}\\
\caption{\label{FigS2}\textbf{Increase of pairwise MI obtained by the MI-IPA: finite-size effect corrections.} In both panels, similar results as those in Fig.~\ref{Fig3}B are reported, but two different finite-size effect corrections to the MI estimates are implemented. Excess pairwise MI of the actual set of correctly paired HK-RRs, compared to the initial random within-species pair assignments and to the final assignment predicted by the MI-IPA, are plotted versus the total number $M$ of sequences in the dataset. (A) Finite-size correction using subsampling of each dataset and fitting versus $1/M$ (see Results). Third-degree polynomial fitting was employed. (B) NSB correction~\cite{Nemenman02,Nemenman04} (see Results). In both panels, naive estimates (see Fig.~\ref{Fig3}B) are also shown for comparison. For each $M$, HK-RR datasets are constructed by picking species randomly from the full dataset, and results are averaged over 50 different such alignments. }
\end{figure}

As with the DCA-IPA~\cite{Bitbol16}, an important potential application of the MI-IPA is to predict new protein-protein interactions between two protein families. A signature of interactions based on exploiting multiple different random initializations of the algorithm was introduced in~\cite{Bitbol16}, and can be applied to the MI-IPA. Briefly, if two given proteins from the families considered are paired by the algorithm for any initial random within-species pairing, it indicates that these two particular proteins may interact. Hence, if many pairs are in this case, it constitutes a clue that proteins from these two families generally interact. In Fig.~\ref{Fig5}, the distributions of the fraction of replicates of the MI-IPA in which each particular pair of proteins is predicted (``replication fraction'') are shown for two interacting pairs of protein families with similar mean number $\langle m_p\rangle$ of pairs per species: the subset of HK-RRs homologous to BASS-BASR, and the homologs of the interacting ABC transporter
proteins MALG-MALK. A pair with no known interaction, composed of homologs of BASR-MALK, is also considered. For both interacting pairs of protein families (Fig.~\ref{Fig5}A-B), the distribution of replication fractions is strongly bimodal: it favors values close to 0 (mostly corresponding to wrong pairs), and close to 1 (mostly corresponding to actual cognate pairs). No such strong bimodality is observed for the pair with no known interaction (Fig.~\ref{Fig5}C), and the observed distribution of replication fractions is closer to the null model constructed by randomly scrambling the amino acids at each site (column) of the alignment, thus removing correlations. Hence, as for the DCA-IPA~\cite{Bitbol16}, the distribution of the replication fraction presents a distinctive shape for interacting protein families. Note that the bimodality observed for the interacting pairs is even stronger with the MI-IPA than with the DCA-IPA, consistent with the better predictive power of the MI-IPA. However, the non-interacting pair was even closer to the null model with the DCA-IPA than it is here.

\begin{figure}[h t b]
\centering
\includegraphics[width=0.55\textwidth]{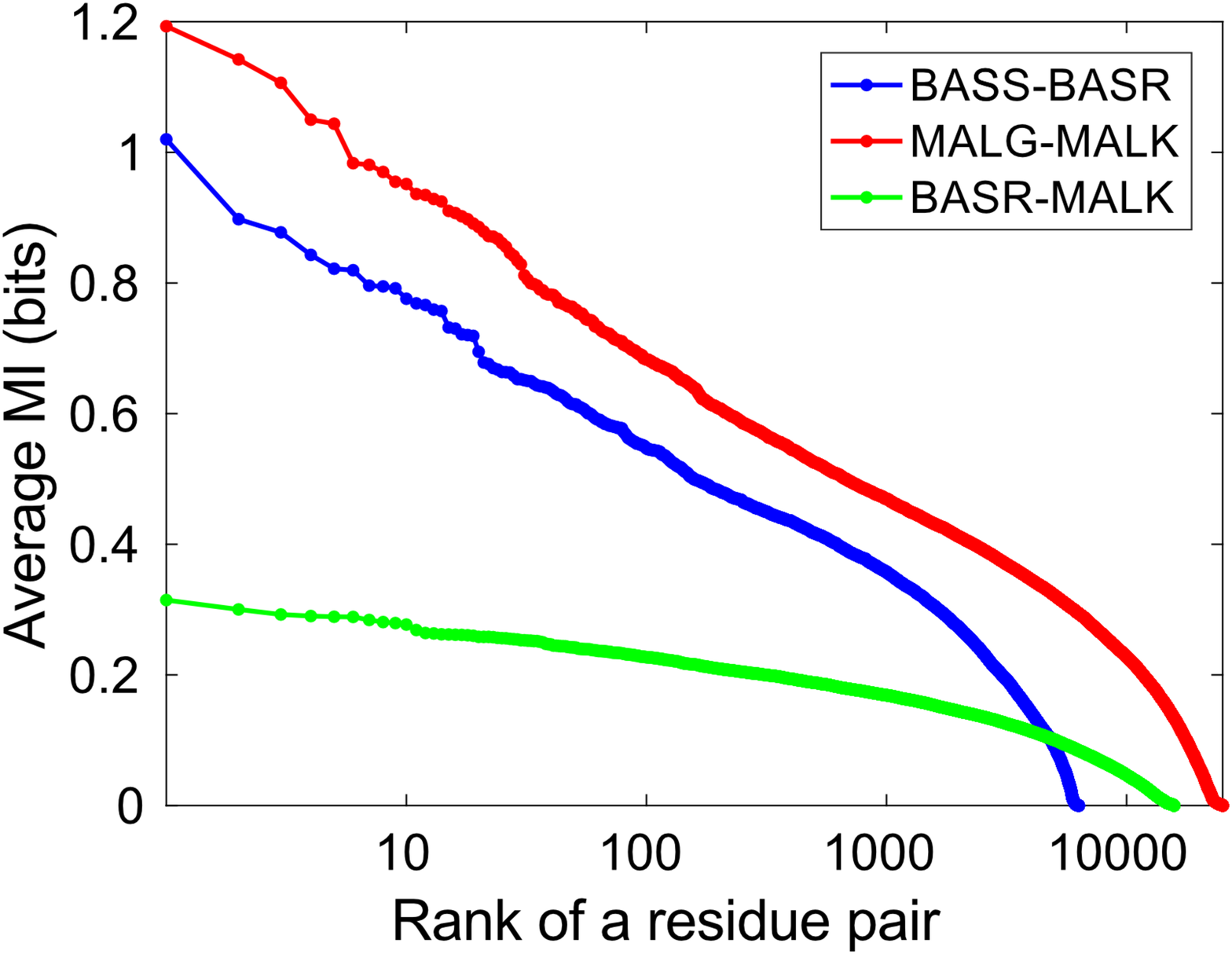}\\
\caption{\label{FigS3}\textbf{Alternative signature of protein-protein interactions. } The MI of each inter-protein pair of amino acids was evaluated at the final iteration of the MI-IPA, for three different pairs of protein families, including two interacting ones (BASS-BASR homologs and MALG-MALK homologs) and one with no known interaction (BASR-MALK homologs). These MI scores were averaged over 500 MI-IPA replicates with $N_\mathrm{increment}=50$ that differ in their initial random pairings, and then ranked by decreasing value. Datasets used are the same as in Fig.~\ref{Fig5}: all alignments include $\sim\!5000$ protein pairs, with average number of pairs per species $\langle m_p\rangle\approx 5$.}
\end{figure}

Another possible approach to identifying interacting protein families is to focus on MI scores, with the intuition that MI between interacting protein families should be higher than those between non-interacting families. DCA-based approaches~\cite{Ovchinnikov14,Feinauer16} to distinguishing interacting protein families from non-interacting ones have employed such strategies, relying on the average value of the DCA interaction scores over the top-ranking inter-protein pairs of residues. If this value is higher than a threshold, the protein families are predicted to interact, and the top-ranking pairs of residues become predicted contacts. In~\cite{Gueudre16}, this strategy was combined with a solution of the paralog-pairing problem. In~\cite{Bitbol16}, a variant based on detecting outliers in DCA interaction scores was proposed and combined with the DCA-IPA. In Fig.~\ref{FigS3}, we show the MI of all inter-protein residue pairs, averaged over 500 replicates of the IPA, and then ranked by decreasing MI value, for two interacting pairs of protein families (BASS-BASR and MALG-MALK, see above) and one without known interactions (BASR-MALK). These plots do not feature strong outliers in the case of interacting protein families, contrary to the DCA interaction scores studied in~\cite{Bitbol16}. However, MI values are substantially larger for the interacting pairs of protein families than for the non-interacting one. This suggests that a score based on MI values, e.g. their average across all inter-protein residue pairs, or across the top-ranked ones, could be employed to distinguish interacting from non-interacting pairs of protein families. In order to predict new protein-protein interactions, one would need to set a threshold on this score, for instance by assessing its performance on known interacting and non-interacting protein families, as for the DCA-based approaches~\cite{Ovchinnikov14,Feinauer16}. 

\begin{figure}[h t b]
\centering
\includegraphics[width=\textwidth]{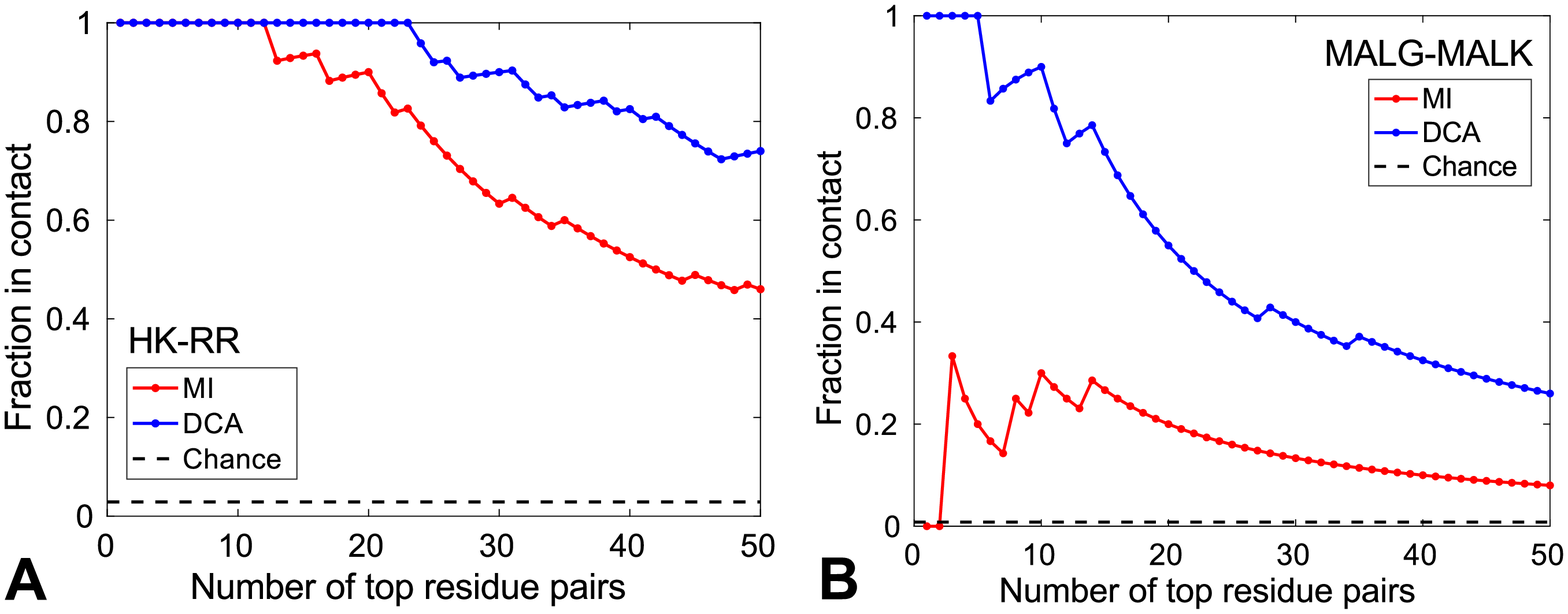}\\
\caption{\label{FigS4}\textbf{Prediction of inter-protein residue contacts. } Fraction among the top inter-protein residue pairs ranked by MI or DCA scores that are in contact in the experimentally-determined three-dimensional structure of complexes. Inter-protein residue pairs were scored both by MI (see Eq.~\ref{MI}) and by the Frobenius norm of the coupling strengths from DCA~\cite{Bitbol16}, and then ranked using these scores. (A) HK-RR dataset; contacts were determined from the experimental structure with PDB identifier 3DGE~\cite{Casino09}. (B) Dataset of homologs of MALG-MALK; contacts were determined from the experimental structure with PDB identifier 3RLF~\cite{Oldham11}. The alignments of $\sim\!5000$ cognate protein pairs employed in the rest of the paper to test the MI-IPA (see Methods) were used. A generous threshold of $8\,\text{\AA}$ on the minimum distance between two amino acids was used to define contacts. In both panels, the chance expectation for finding contacts (i.e. the overall fraction of inter-protein residue pairs that are in contact) is shown (dashed lines).}
\end{figure}

The absence of outliers in MI in Fig.~\ref{FigS3} for the interacting pairs of protein families likely stems from the fact that interactions between amino acids can cause indirect correlations, which are disentangled by DCA but not by MI. Indeed, the outliers in DCA scores found in~\cite{Bitbol16} mainly correspond to contacting residue pairs. DCA performs better than MI at identifying contacting residue pairs~\cite{Weigt09,Marks11,Morcos11}, and the top-ranking MI residue pairs are less likely to be actual contacts than the top-ranking DCA residue pairs (see Fig.~\ref{FigS4}). Note that the performance of both DCA and MI at uncovering interacting residue pairs can often be improved by the average product correction (APC)~\cite{Dunn08,Jones12,Ekeberg13}. However, this correction did not substantially improve inter-protein contact prediction in HK-RRs or yield strong outliers in MI, so we did not employ it. 

Given the success of the MI-IPA at providing global signatures of interactions between two protein families and at globally pairing interacting paralogs, we can also ask whether MI-based methods can accurately predict partners of isolated proteins. DCA-based methods have been used to predict interacting partners of ``orphan'' HKs and RRs involved in prokaryotic two-component signaling systems~\cite{Procaccini11,Baldassi14}. These particular proteins are not encoded on the genome close to their cognate partners, contrary to the vast majority of HKs and RRs. Crosstalk between non-cognate HKs and RRs, which exists in a few cases~\cite{Laub07}, has also been studied using DCA~\cite{Procaccini11,Cheng16}. Given the striking success of the MI-IPA at identifying cognate interaction partners among HK-RRs, we asked whether MI-based scores can predict orphan partnership and crosstalk. Table~\ref{Table} reports the performance of MI and DCA-based scores at predicting known orphan and crosstalk partners using only sequences. Specifically, PMI scores (see Eq.~\ref{PMI}) and DCA-based scores (direct couplings~\cite{Bitbol16}) were calculated from the full alignment of cognate HK-RR pairs (which excludes orphans, see Methods). Next, new alignments including both cognate and orphan proteins, and RRs paired with non-HisKA HKs (see Methods), were considered, focusing on species where orphan and crosstalk partners were reported in the literature, either from \textit{in vivo} or from \textit{in vitro} experimental assays. For each HK considered, all RRs in the species were ranked using either MI-based scores (see Eq.~\ref{SAB}) or DCA-based scores (effective interaction energies~\cite{Bitbol16}) for partnership with this HK. With both methods, known partners often ranked among the top potential partners, showing that DCA and MI capture well these interactions. Besides, we checked that our DCA-based predictions were similar to previous ones reported in~\cite{Procaccini11} (see Table~\ref{Table}). Out of the 27 known orphan and/or crosstalk pairs we considered, 7 obtained the same rank among all possible partners with MI as with DCA, 15 a lower rank, and 5 a higher rank (see Table~\ref{Table}). Hence, our MI-based approach performs quite well at predicting orphan and crosstalk partners, but slightly less well than DCA.

\section*{Discussion}

Here, we introduced a method based on MI to predict interacting partners among the paralogs of two protein families, starting from sequences only. Specifically, our iterative pairing algorithm (MI-IPA) finds an assignment of protein pairs that approximately maximizes the MI between the sequences of the two protein families. We demonstrated that the MI-IPA allows one to predict interacting protein pairs with high accuracy, starting from sequence data only. High performance is obtained even in the absence of an initial training set of known interacting pairs. We also showed how the MI-IPA could be employed to discover new protein-protein interactions from sequence data.

In~\cite{Bitbol16}, us and colleagues introduced a similar iterative pairing algorithm (DCA-IPA) that approximately maximized an effective interaction energy between proteins, instead of MI. This effective interaction energy was calculated from a global statistical model, more specifically a pairwise maximum entropy model, approximately inferred from the empirical one and two-body frequencies of the sequence data. Such models, often called Direct Coupling Analysis (DCA), have been successful at predicting amino-acid pairs that are in contact in folded proteins~\cite{Weigt09,Marks11,Morcos11}, and have permitted prediction of the three-dimensional structure of proteins from sequence data~\cite{Weigt09,Marks11,Morcos11,Sulkowska12,Ovchinnikov14,Ovchinnikov15}. For structure prediction, these global statistical models outperform the use of MI~\cite{Weigt09,Marks11}, and have yielded major progress in the field, despite the promising results obtained by MI-based methods implementing corrections for background MI~\cite{Dunn08}. In the specific case of interacting proteins, DCA allows one to simultaneously infer interaction partners and structural contacts between them~\cite{Gueudre16}.

Here, we demonstrated that the MI-IPA performs at least as well as the DCA-IPA, and often outperforms it. The MI-IPA is also faster (see Methods) and requires fewer iterations to reach good performance. Our results highlight an interesting difference between the prediction of contacting pairs of residues, where DCA substantially outperforms MI, and the prediction of interacting partners, where MI often outperforms DCA. Crucially, we have achieved an accurate identification of interacting partners without the need to build a global statistical model of the sequence data highlighting effective pairwise interactions between contacting residues, which is at the root of DCA. An important motivation underlying DCA and other related approaches, such as the Bayesian network method of~\cite{Burger08} and the sparse inverse covariance estimation of~\cite{Jones12}, is to disentangle correlations arising caused by direct interactions from indirect correlations due to a chain of couplings~\cite{Lapedes99,Weigt09,Burger08,Jones12}.  While this distinction between direct and indirect correlations is crucial to infer which residues are in contact in the folded protein, its importance is probably reduced in the determination of interacting partners among paralogous proteins. Besides, MI has been extremely successful in determining ``specificity residues'' crucial in the interactions between HKs and their cognate RRs: these residues are those involved in interprotein residue pairs with highest MIs~\cite{Skerker08,Podgornaia13,Podgornaia15}. Strikingly, mutating only these few specificity residues has allowed to fully switch the specificity of HK-RR pairs~\cite{Skerker08}. In addition, MI has revealed a cluster of co-evolving dynamic residues that are not in direct contact, but that are important to interaction specificity, through their involvement in the conformational arrangement of the active site residues both in the HK and in the RR, as confirmed by mutagenesis and NMR studies~\cite{Szurmant08}. Hence, our present results reinforce the conclusion of these earlier studies, showing that MI can accurately reveal the specificity of interacting protein pairs. 

Finally, the MI between two sequence sites simply measures how much observing a residue at one site tells us about the other. It is entirely agnostic regarding the origin of this statistical dependence, and incorporates contributions from phylogeny~\cite{Casari95,Qin18} as well as from those arising from functional selection. In addition, the latter contributions include those arising directly or indirectly from structural contacts, but might also involve aspects of protein function other than structural ones, including collective correlations between residues~\cite{Halabi09,Rivoire16,Yan17,Cunningham17,WangXX}. This stands in contrast with DCA, which lays the emphasis on interactions between residues that are in direct contact in the three-dimensional protein structure. The fact that the MI-IPA slightly outperforms the DCA-IPA thus constitutes a hint that sources of covariation other than those that maintain structural contacts help to identify interaction partners among paralogs. This might be due to multiple effects. In particular, global functional selection could potentially affect both interacting partners together. For instance, a functionally important mechanical deformation of the complex formed by these partners could be subject to selection, yielding collective correlations that extend in the sequences of both partners~\cite{Yan17,WangXX}. Besides, interacting partners may also share more phylogenetic history than non-interacting proteins, e.g. if the genes encoding the partners tend to be duplicated and/or horizontally transferred together~\cite{Alm06,Capra12,Rowland14}. Then, phylogenetic correlations could aid the prediction of interacting partners among paralogous proteins, despite being deleterious to residue contact identification~\cite{Qin18}. Next, it will be interesting to further investigate these various sources of covariation, both functional and phylogenetic. This should be useful for the particular problem of prediction of interacting protein pairs, as well as for the more general understanding of the sequence-function relationship in proteins.

\section*{Methods}

Here, we first explain the different steps of the MI-IPA and briefly discuss its run time. Next, we describe the datasets used and the way the statistics of these alignments are computed. Matlab implementations of both the MI-IPA and the DCA-IPA on our standard HK-RR dataset are freely available at \url{https://doi.org/10.5281/zenodo.1421781} and\\ \url{https://doi.org/10.5281/zenodo.1421861} respectively. The various sequence datasets discussed here are available as S1 Data at \url{https://doi.org/10.1371/journal.pcbi.1006401}.

\subsection*{Iterative pairing algorithm based on mutual information (MI-IPA)}

Ref.~\cite{Bitbol16} introduced an iterative pairing algorithm (IPA) to predict interaction partners among paralogs from two protein families. It essentially performs an approximate maximization of the average effective interaction energy between pairs of proteins comprised of one protein of family $\mathcal{A}$ and one of family $\mathcal{B}$, and the effective interaction energy is calculated from a pairwise maximum entropy model (see also~\cite{Gueudre16}). Here, we propose the MI-IPA, a variant of the IPA that approximately maximizes MI via $S_\mathrm{X}$ (see Eq.~\ref{Sx1}), instead of the effective interaction energy. Importantly, the MI-IPA does not require the construction of a global statistical model of the data, contrary to the algorithms from Refs.~\cite{Bitbol16,Gueudre16}. Let us now describe each step of an iteration of the MI-IPA, after explaining how the CA is constructed for the very first iteration. 

For simplicity, we assume that in each species, there is the same number of proteins of family $\mathcal{A}$ and of family $\mathcal{B}$. If this is not the case, an injective matching strategy can be used~\cite{Gueudre16}, so in each species, the relevant number of proteins is the minimum of the number of proteins of family $\mathcal{A}$ and of proteins of family $\mathcal{B}$. 

\subsubsection*{Initialization of the CA}
If starting from a training set of known interaction partners AB, the CA for the first iteration of the IPA is built from the pairs AB in this training set. In subsequent iterations, the training set pairs are always kept in the CA, and additional pairs with the highest confidence scores (see below) are added to the CA.

In the absence of a training set, each protein A of the dataset where we wish to predict pairings is randomly paired with a protein B from its species. All these random pairs are included in the CA for the first iteration of the MI-IPA. Hence, this initial CA contains a mixture of correct and incorrect pairs, with one correct pair per species on average. At the second iteration, the CA is built using only the $N_\mathrm{increment}$ AB pairs with the highest confidence scores obtained from this first iteration. 

Now that we have explained the initial construction of the CA, let us describe each step of an iteration of the MI-IPA.

\subsubsection*{1. Calculation of pointwise mutual informations (PMI)}
Each iteration begins by the calculation of PMI scores from the CA of paired AB sequences. The empirical one- and two-site frequencies, $f_i(\alpha)$ and $f_{ij}(\alpha,\beta)$, of occurrence of amino-acid states $\alpha$ (or $\beta$) at each site $i$ (or $j$) are computed for the CA, using a specific weighting of similar sequences, and a pseudocount correction (see below)~\cite{Weigt09,Procaccini11,Marks11,Morcos11}. The PMI of each residue pair $(\alpha,\beta)$ at each pair of sites $(i,j)$, defined in Eq.~\ref{PMI}, is then estimated from these frequencies as
\begin{equation}
\textrm{PMI}_{ij}(\alpha,\beta)=\log\left[\frac{f_{ij}(\alpha,\beta)}{f_i(\alpha)f_j(\beta)}\right]\,.
\label{PMIf}
\end{equation}

\subsubsection*{2. Calculation of pairing scores for all possible AB pairs}
Having calculated PMIs on the CA, we next turn to the dataset where we wish to predict pairings. The MI-based pairing score $S_\mathrm{AB}$ of each possible AB pair within each species in the dataset is calculated by summing all the inter-protein PMIs involved, as defined in Eq.~\ref{SAB}.

\subsubsection*{3. AB pair assignments and ranking by confidence score}
We make one-to-one AB pairs within each species in the dataset by maximizing the sum of the scores of all pairs in this species. Considering the matrix of the scores of all possible pairs, this amounts to choosing one element per line and per column, such that the sum of all of them is maximal. This assignment problem is solved exactly and efficiently using the Hungarian algorithm (also known as the Munkres algorithm)~\cite{Kuhn55,Munkres57,HungAlg}. Each pair is scored by a confidence score $\Delta S_\mathrm{AB}$, which is the difference between the sum of the scores of the assigned pairs in the species and that obtained by using the Hungarian algorithm again while this pair is disallowed. Once pairs are made and confidence scores are calculated in each species, all the assigned AB pairs from all species are ranked in order of decreasing confidence score. 

\begin{figure}[h t b]
\centering
\includegraphics[width=0.5\textwidth]{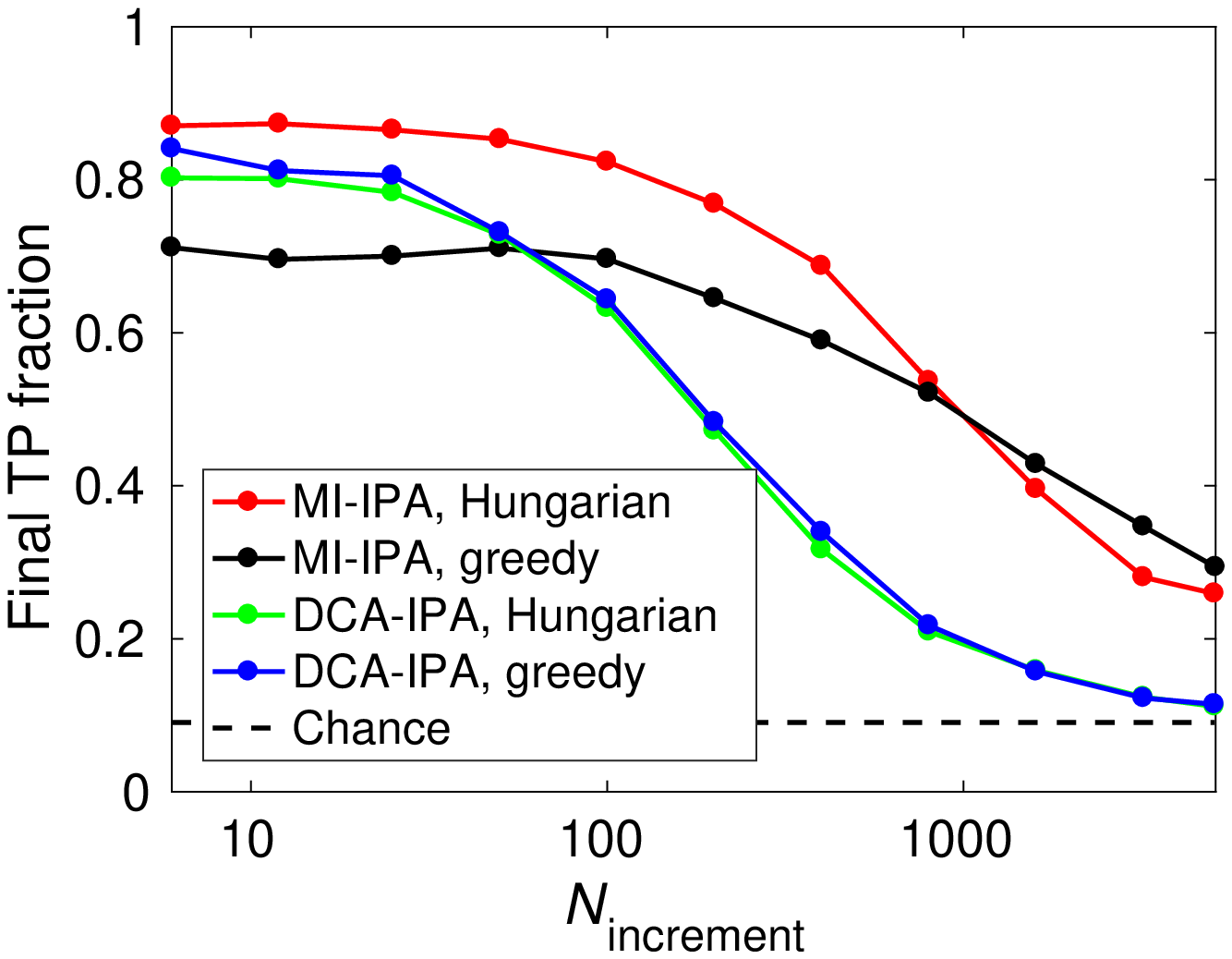}\\
\caption{\label{FigS5}\textbf{IPA assignment variants.} Final TP fraction obtained versus $N_{\mathrm{increment}}$ for the standard HK-RR dataset, starting from random within-species HK-RR pairs. For both the MI-IPA and the DCA-IPA, two variants of the protein pair assignment strategy are presented. Once pairing scores are computed, pairs are assigned within each species either using the Hungarian algorithm, which maximizes the sum of scores in the species, or using a greedy algorithm, that favors individual pairs with top scores (see Methods). In all the rest of the paper, the Hungarian algorithm is used for the MI-IPA, and the greedy one is used for the DCA-IPA, as in ~\cite{Bitbol16}. All results are averaged over 50 replicates employing different initial random pairings, and the dashed line represents the average TP fraction obtained for random within-species HK-RR pairings.}
\end{figure}

Note that in~\cite{Bitbol16}, we had used an approximate greedy approach to the assignment problem, where the pair with lowest energy is selected first, and the two proteins involved are removed from further consideration, and the process is repeated until all As and Bs are paired. For the DCA-IPA of~\cite{Bitbol16}, the greedy algorithm performed marginally better than the Hungarian algorithm, while in the present MI-IPA, the Hungarian algorithm yields better performance (see Fig.~\ref{FigS5}). This slight difference may be explained by the fact the PMI scores are most meaningful collectively, as their sum tends to a pairwise approximation of MI (see Eq.~\ref{MIlim}). 

\subsubsection*{4. Update of the CA}

The ranking of the AB pairs is used to pick those pairs that are included in the CA at the next iteration. Pairs with a high confidence score are more likely to be correct because there was less ambiguity in the assignment. The number of pairs in the CA is increased by $N_\mathrm{increment}$ at each iteration, and the MI-IPA is run until all the As and Bs in the dataset have been paired and added to the CA. In the last iteration, all pairs assigned at the second to last iteration are included in the CA. 

If starting from a training set of AB pairs, the $N_\mathrm{start}$ training pairs remain in the CA throughout. The As and Bs from all the other pairs in the CA are re-paired and re-scored at each iteration, and only re-enter the CA if their confidence score is sufficiently high. In other words, at the first iteration, the CA only contains the $N_\mathrm{start}$ training pairs. Then, for any iteration number $n>1$, it contains these exact same $N_\mathrm{start}$ training pairs, plus the $(n-1)N_\mathrm{increment}$ assigned AB pairs that had the highest confidence scores at iteration number $n-1$. 

In the absence of a training set, all As and Bs in the dataset are paired and scored at each iteration, and all the pairs of the CA are fully re-picked at each iteration based on the confidence score. For any iteration number $n>1$, the CA contains the $(n-1) N_\mathrm{increment}$ assigned AB pairs that had the highest confidence scores at iteration number $n-1$. 

Once the new CA is constructed, the next iteration can start.

\subsection*{Run time}
The run time of the MI-IPA strongly depends on the number of iterations made, as well as on dataset size (length of sequences, number of sequences in the dataset). Because the inversion of the correlation matrix is not necessary, the MI-IPA is substantially faster than the DCA-IPA, and this difference is stronger for longer sequences. In practice, for our standard HK-RR dataset, one round of calculation of all PMI scores from sequence data (Eq.~\ref{PMI}) takes 7s, while for DCA, the calculation of all pairwise couplings takes 9s. The longest concatenated sequences considered here were XDHA-XDHC (see Fig.~\ref{Fig4}), with total length $L=911$ (to be compared to 176 for HK-RRs): in this case, the calculation of all PMI scores takes 135s, while for DCA, the calculation of pairwise couplings takes 289s. Times are given for a single processor (Intel Core i7) of a standard laptop computer. Regarding the full MI-IPA, run times with all standard datasets used here were under 10 hours.

\subsection*{Comparison to the DCA-IPA}
Throughout, the DCA-IPA was used as described in~\cite{Bitbol16}, with the same parameters (threshold Hamming distance for weighting neighboring sequences $\theta=0.3$, pseudocount weight $\Lambda=0.5$).

\subsection*{Dataset construction}

We use the HK-RR datasets described in~\cite{Bitbol16}. Briefly, the complete dataset was built using the online database P2CS~\cite{Barakat09,Ortet15}, yielding a total of 23,424 HK-RR pairs (known by genome proximity) from 2102 different species. We focused on the protein domains through which HKs and RRs interact, which are the Pfam HisKA domain present in most HKs (64 amino acids) and the Pfam Response\_reg domain present in all RRs (112 amino acids). 

In most of the paper, and as in~\cite{Bitbol16}, we focused on a smaller ``standard dataset'' extracted from this complete dataset, both because protein families that possess as many members as the HKs and RRs are atypical, and in view of computational time constraints. This standard dataset was constructed by picking species randomly, and comprises 5064 pairs from 459 species. In our datasets, we discarded sequences from species that contain only one pair, for which pairing is unambiguous. It allows us to assess the impact of training set size ($N_\mathrm{start}$) without including an implicit training set via these pairs, and it also enables us to address prediction in the absence of any known pairs (no training set). 

While we used HK-RRs as the main benchmark for the MI-IPA, we assessed the generality of its performance by applying it to several other pairs of protein families. For these proteins, paired alignments of homologs of known \textit{Escherichia coli} interacting protein pairs were built using a method adapted from~\cite{Ovchinnikov14}, as detailed in~\cite{Bitbol16}. Note that we kept full sequences, without restricting them to Pfam domains, for PAAH-PAAJ and XDHC-XDHA.

\subsection*{Statistics of the concatenated alignment (CA)}

Let us consider a CA of paired AB sequences. At each site $i\in\{1,..,L\}$, where $L$ is the length of an AB sequence, a given concatenated sequence can feature any of 21 amino acid states $\alpha$.

\begin{figure}[h t b]
\centering
\includegraphics[width=\textwidth]{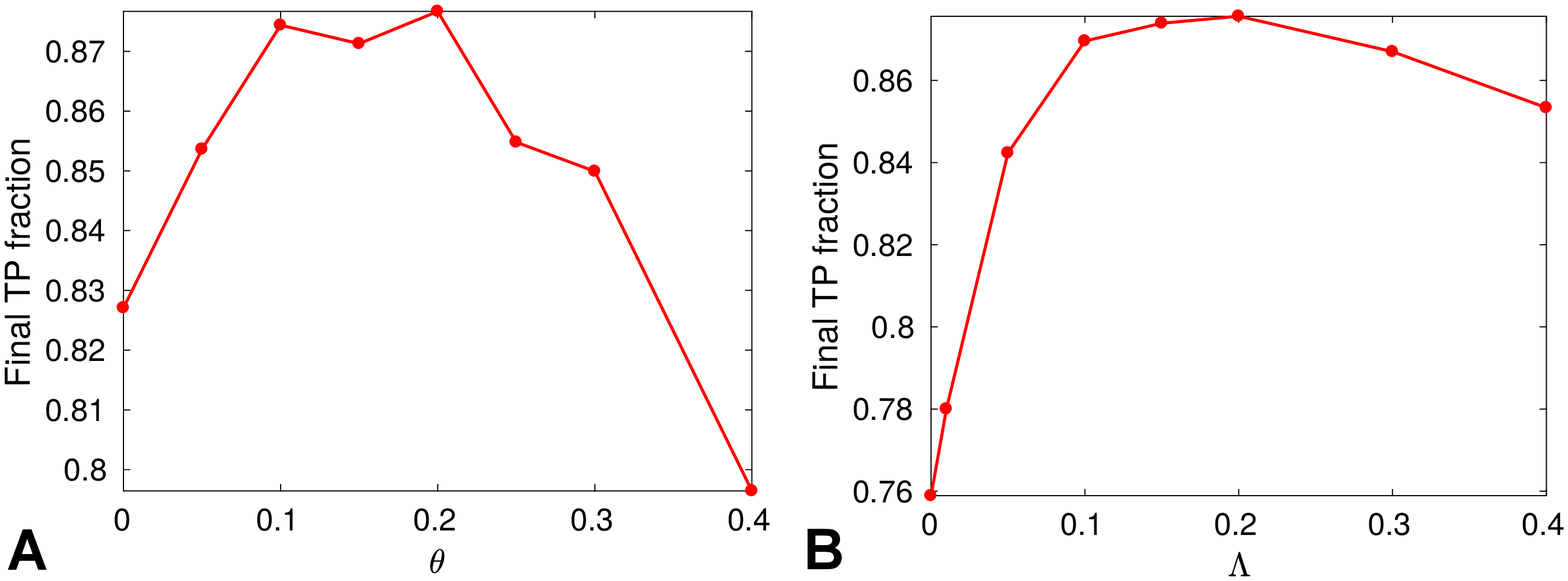}\\
\caption{\label{FigS6}\textbf{Impact of the weighting of similar sequences and of the pseudocount.} (A) Final TP fraction obtained for the standard HK-RR dataset, starting from random within-species HK-RR pairs, with $N_{\mathrm{increment}}=6$, for various values of the threshold $\theta$ of mean Hamming distance per site below which two sequences are considered as neighbors and weighted accordingly (see Methods). Pseudocount $\Lambda=0.15$ was used. (B) Similar graph, but varying the pseudocount $\Lambda$ (see Methods), at fixed $\theta=0.15$. In both panels, all results are averaged over 50 replicates employing different initial random pairings.}
\end{figure}

The raw empirical frequencies $f_i(\alpha)$ and $f_{ij}(\alpha,\beta)$, obtained by counting the sequences where given residues occur at given sites and dividing by the number $M$ of sequences in the CA, are subject to sampling bias, due to phylogeny and to the choice of species that are sequenced~\cite{Morcos11,Marks11}. Hence, we use a standard correction that re-weights ``neighboring'' concatenated sequences with mean Hamming distance per site $<\theta$. In practice $\theta=0.15$ was found to yield the best results, but the dependence of performance on $\theta$ is weak (see Fig.~\ref{FigS6}A). The weight associated to a given concatenated sequence $a$ is $1/m_a$, where $m_a$ is the number of neighbors of $a$ within the threshold (including the sequence considered)~\cite{Marks11,Procaccini11,Morcos11}. This allows one to define an effective total number of sequences, $M_\textrm{eff}=\sum_{a=1}^M 1/m_a$.

We also introduce pseudocounts via a parameter $\Lambda$~\cite{Weigt09,Procaccini11,Marks11,Morcos11} to avoid issues due e.g. to amino-acid pairs that never appear. Indeed, those can yield mathematical difficulties, such as diverging PMI estimates. Note that pseudocounts are widely used in DCA too~\cite{Morcos11}. The corrected one-body frequencies $\tilde{f}_i$ read
\begin{equation}
\tilde{f}_i(\alpha)=\frac{\Lambda}{q}+(1-\Lambda)f_i(\alpha)\,,
\label{fi}
\end{equation}
where $q=21$ is the number of possible states per site. Similarly, the corrected two-body frequencies $\tilde{f}_{ij}$ read
\begin{align}
\tilde{f}_{ij}(\alpha,\beta)&=\frac{\Lambda}{q^2}+(1-\Lambda)f_{ij}(\alpha,\beta)\textrm{ if }i\neq j\,, \label{fij}\\
\tilde{f}_{ii}(\alpha,\beta)&=\frac{\Lambda}{q} \delta_{\alpha\beta}+(1-\Lambda)f_{ii}(\alpha,\beta)= \tilde{f}_i(\alpha)\delta_{\alpha\beta}\,, \label{fii}
\end{align}
 where $\delta_{\alpha\beta}=1$ if $\alpha=\beta$ and 0 otherwise. We found that $\Lambda=0.15$ yields the best performance of the MI-IPA, but the dependence of performance on $\Lambda$ is weak (see Fig.~\ref{FigS6}B). Note that this value is lower than the typical value used in DCA ($\Lambda=0.5$)~\cite{Morcos11,Marks11}.

Note that our demonstration that the average pairing score $S_\mathrm{X}$ of an assignment tends to the sum of MIs of inter-protein residue pairs (see Eqs.~\ref{Sx1} and~\ref{MIlim}) did not include proximity weightings and pseudocounts. In practice, at each iteration of our algorithm, assignments are made within each species separately, a scale at which it is convenient to just use the sum of pairing scores $S_\mathrm{AB}$ (see Eq.~\ref{SAB}), and find the assignment that maximizes it.  This means that the convergence to the sum of MIs is approximate when using proximity weightings and pseudocounts to calculate the frequencies in the CA and estimate PMIs. 

Note also that empirical frequencies without weightings or pseudocounts were employed to study how well the MI-IPA maximizes MI (Figs.~\ref{Fig3} and~\ref{FigS2}).

\section*{Acknowledgments}
A.-F. B. thanks Ned S. Wingreen for illuminating discussions. A.-F. B. acknowledges the Aspen Center for Physics, which is supported by NSF Grant PHY-1607611, and more particularly the participants of the 2017 Aspen Working Group on ``Covariance Analysis of Proteins'' for stimulating discussions.

\beginsupplement

\newpage

\section*{Supplementary figures and tables}

\begin{table}[htb]
    \centering
    \begin{tiny}
    
\begin{tabular}{| l | p{1.2cm} | p{1.9cm}|p{1.1cm} | p{1.9cm} | l | p{0.8cm} | p{0.4cm}| p{0.4cm} | p{1.2cm} |}
\hline
Species & HK usual name & HK ordered locus name & RR usual name & RR ordered locus name & Ref. & Number of RRs & MI rank & DCA rank & Comparison with~\cite{Procaccini11}\\ 
\hline
\textit{Thermotoga maritima} &  & TM0853 &  & TM0468 & \cite{Casino07} & 11 & 2 & 2 & \\
\textit{Synechococcus elongatus} & SasA/CikA & Synpcc7942\_2114 & SrrA & Synpcc7942\_2416 & \cite{Kato12} & 21 & 2 & 2 & \\  
\textit{Synechococcus elongatus} & SasA/CikA & Synpcc7942\_2114 & RpaB & Synpcc7942\_1453 & \cite{Kato12} & 21 & 6 & 4 & \\ 
\textit{Synechococcus elongatus} & SasA/CikA & Synpcc7942\_2114 & RpaA & Synpcc7942\_0095 & \cite{Kato12,Gutu13} & 21 & 7 & 5 & \\ 
\textit{Synechococcus elongatus} &  & Synpcc7942\_0453 & Ycf29 & Synpcc7942\_1860 & \cite{Kato12} & 21 & 2 & 12 & \\ 
\textit{Synechococcus elongatus} &  & Synpcc7942\_2242 & SrrB & Synpcc7942\_0556 & \cite{Kato12} & 21 & 8 & 5 & \\ 
\textit{Synechococcus elongatus} & NblS & Synpcc7942\_0924 & RpaB & Synpcc7942\_1453 & \cite{Kato12} & 21 & 1 & 1 & \\ 
\textit{Synechococcus elongatus} & NblS & Synpcc7942\_0924 & SrrA & Synpcc7942\_2416 & \cite{Kato12} & 21 & 3 & 4 & \\ 
\textit{Synechocystis} & Hik2 & slr1147 & Rre1 & slr1783 & \cite{Kato12, Vidal15} & 41 & 2 & 7 & \\ 
\textit{Myxococcus xanthus} & CrdS & MXAN\_5184 & CrdA & MXAN\_5153 & \cite{Willett13} & 138 & 1 & 1 & \\ 
\textit{Myxococcus xanthus} & SasS & MXAN\_1249 & SasR & MXAN\_1245 & \cite{Willett13} & 138 & 9 & 3 & \\ 
\textit{Caulobacter crescentus} & DivJ & CC\_1063 & DivK  & CC\_2463 & \cite{Skerker05} & 46 & 8 & 1 & Same\\ 
\textit{Caulobacter crescentus} & DivJ & CC\_1063 & PleD & CC\_2462 & \cite{Skerker05} & 46 & 13 & 2 & Same\\ 
\textit{Caulobacter crescentus} & PleC & CC\_2482 & PleD & CC\_2462 & \cite{Skerker05} & 46 & 2 & 1 & Better\\ 
\textit{Caulobacter crescentus} & PleC & CC\_2482 & DivK & CC\_2463 & \cite{Skerker05} & 46 & 3 & 4 & Worse*\\ 
\textit{Caulobacter crescentus} & CenK & CC\_0530 & CenR & CC\_3743 & \cite{Skerker05, Procaccini11} & 46 & 17 & 7 & Better\\ 
\textit{Caulobacter crescentus} & DivL & CC\_3484 & DivK & CC\_2463 & \cite{Procaccini11} & 46 & 16 & 24 & Worse\\ 
\textit{Caulobacter crescentus} &  & CC\_1062 & DivK & CC\_2463 & \cite{Procaccini11} & 46 & 14 & 2 & Worse**\\ 
\textit{Clostridium acetobutylicum} &  & CA\_C0903 & Spo0A & CA\_C2071 & \cite{Steiner11} & 42 & 22 & 12 & \\ 
\textit{Clostridium acetobutylicum} &  & CA\_C3319 & Spo0A & CA\_C2071 & \cite{Steiner11} & 42 & 22 & 12 & \\ 
\textit{Bacillus subtilis} & KinA & BSU13990 & Spo0F & BSU37130 & \cite{Procaccini11} & 35 & 1 & 1 & Same\\ 
\textit{Bacillus subtilis} & KinB & BSU31450 & Spo0F & BSU37130 & \cite{Procaccini11} & 35 & 6 & 3 & Better\\ 
\textit{Bacillus subtilis} & KinC & BSU14490 & Spo0F & BSU37130 & \cite{Procaccini11} & 35 & 6 & 1 & Same\\ 
\textit{Bacillus subtilis} & KinD & BSU13660 & Spo0F & BSU37130 & \cite{Procaccini11} & 35 & 1 & 1 & Same\\ 
\textit{Bacillus subtilis} & KinE & BSU13530 & Spo0F & BSU37130 & \cite{Procaccini11} & 35 & 2 & 1 & Same\\ 
\textit{Streptococcus mutans} (pair) &  VicK & Smu\_1516 & VicR & Smu\_1517 & \cite{Downey14} & 15 & 1 & 1 & \\ 
\textit{Streptococcus mutans} &  VicK & Smu\_1516 & GcrR & Smu\_1924 & \cite{Downey14} & 15 & 2 & 2 & \\
\hline
\end{tabular}
  
\end{tiny}
\caption{   \textbf{Orphan partnership and crosstalk prediction in HK-RRs by MI and DCA.} MI and DCA scores trained on the full alignment of cognate HK-RR pairs (excluding orphans, see Methods) were used to score all possible HK-RR pairs, including orphans, in species where orphan and crosstalk partners have been reported in the literature. For each HK considered, RRs were ranked using the MI and DCA scores. The table indicates the species, the HK and RR names, the literature references where partnership was reported, the total number of RRs in the species considered, and the rank among all these RRs of the RR considered, both using MI scores and using DCA scores. Finally, we report a comparison of the DCA ranking with the previous one in~\cite{Procaccini11}: does our ranking give a better (higher) or a worse (lower) rank to the known partner than this previous ranking? Special mentions in the table: (pair): the cognate pair was included, in order to compare it with the crosstalk partner; Worse*: 2 ranks lower (including an inversion between PleD and DivK, which had close scores in~\cite{Procaccini11}); Worse**: 1 rank lower (including an inversion between PleD and DivK, which had close scores in~\cite{Procaccini11}).  }
      \label{Table}
  \end{table}

\end{document}